\documentclass[aps,pra,showpacs,superscriptaddress,twocolumn]{revtex4-1} 
\usepackage{amsmath,amsfonts,amssymb,amsthm}
\usepackage{mathrsfs}
\usepackage{graphicx,rotating,booktabs}
\usepackage{epsfig} 
\usepackage{setspace}
\usepackage{array}
\usepackage{braket} 
\usepackage{bbm} 
\usepackage{xcolor} 
\usepackage{verbatim}
\usepackage{times} 
\usepackage{mathtools} 
\usepackage{placeins} 
\usepackage{siunitx}
\usepackage{tensor}
\usepackage{bm}
\usepackage{scalerel,stackengine}
\RequirePackage{color} 
\RequirePackage[colorlinks=true, urlcolor=blue, anchorcolor=blue, citecolor=blue, filecolor=blue, linkcolor=blue, menucolor=blue, linktocpage=true, pdfproducer=medialab, pdfa=true]{hyperref} 
\DeclareFontFamily{OT1}{pzc}{}
\DeclareFontShape{OT1}{pzc}{m}{it}{<-> s * [1.10] pzcmi7t}{}
\DeclareMathAlphabet{\mathpzc}{OT1}{pzc}{m}{it}
\DeclareFontFamily{U}{BOONDOX-calo}{\skewchar\font=45 }
\DeclareFontShape{U}{BOONDOX-calo}{m}{n}{
 <-> s*[1.05] BOONDOX-r-calo}{}
\DeclareFontShape{U}{BOONDOX-calo}{b}{n}{
 <-> s*[1.05] BOONDOX-b-calo}{}
\DeclareMathAlphabet{\mathcalboondox}{U}{BOONDOX-calo}{m}{n}
\SetMathAlphabet{\mathcalboondox}{bold}{U}{BOONDOX-calo}{b}{n}
\DeclareMathAlphabet{\mathbcalboondox}{U}{BOONDOX-calo}{b}{n}
\DeclareFontFamily{T1}{calligra}{}
\DeclareFontShape{T1}{calligra}{m}{n}{<->s*[1.44]callig15}{}
\DeclareMathAlphabet\mathcalligra   {T1}{calligra} {m} {n}
\DeclareMathAlphabet\euscr      {U}{rsfso}{m}{n}
\DeclareMathAlphabet\mathzapf       {T1}{pzc} {mb} {it}
\DeclareMathAlphabet\urwscr{U}{urwchancal}{m}{n}
\DeclareMathAlphabet\rsfscr{U}{rsfso}{m}{n}
\DeclareMathAlphabet\euscr{U}{eus}{m}{n}
\DeclareFontEncoding{LS2}{}{}
\DeclareFontSubstitution{LS2}{stix}{m}{n}
\DeclareMathAlphabet\stixcal{LS2}{stixcal}{m} {n}
\def\apeqA{\SavedStyle\sim}
\def\apeq{\setstackgap{L}{\dimexpr.5pt+1.5\LMpt}\ensurestackMath{%
  \ThisStyle{\mathrel{\Centerstack{{\apeqA} {\apeqA}}}}}}
\def\apeqA{\SavedStyle\sim}
\def\apeq{\setstackgap{L}{\dimexpr.5pt+1.5\LMpt}\ensurestackMath{%
 \ThisStyle{\mathrel{\Centerstack{{\apeqA} {\apeqA}}}}}}
\newcommand{\eqd}{\vcentcolon=}

\newcommand{\abs}[1]{\left\lvert#1\right\rvert}
\DeclareMathOperator{\tr}{tr}
\newtheorem{defin}{DEFINITION}
\newtheorem{prop}{PROPOSITION}

\newtheorem{lem}{LEMMA}

\newcommand\mydots{\hbox to 1em{.\hss.\hss.}}
\newcommand\smalldots{\hbox to 0.7em{.\hss.\hss.}}
\usepackage{array}
\newcommand{\be}{\begin{equation}}
\newcommand{\ee}{\end{equation}}
\newcommand{\ben}{\begin{equation*}}
\newcommand{\een}{\end{equation*}}
	
\begin{document} 
\title{Estimation of general Hamiltonian parameters via controlled energy measurements} 
\author{Luigi Seveso}
\email{luigi.seveso@unimi.it} 
\affiliation{Quantum Technology Lab, Dipartimento di Fisica dell'Universit\`a degli Studi di Milano, I-20133 Milano, Italia} 
\author{Matteo G. A. Paris}
\email{matteo.paris@fisica.unimi.it} 
\affiliation{Quantum Technology Lab, Dipartimento di Fisica dell'Universit\`a degli Studi di Milano, I-20133 Milano, Italia}
\affiliation{
Istituto Nazionale di Fisica Nucleare, Sezione di Milano, I-20133 Milano, Italy}
\begin{abstract} 
The quantum Cram\'er-Rao theorem states that the quantum Fisher information (QFI) bounds the best achievable precision in the estimation of a quantum parameter $\xi$. This is true, however, under the assumption that the measurement employed to extract information on $\xi$ are regular, i.e.~neither its sample space nor its positive-operator valued elements depend on the (true) value of the parameter. A better performance may be achieved by relaxing this assumption. In the case of a general Hamiltonian parameter, i.e.~when the parameter enters the system's Hamiltonian in a non-linear way (making the energy eigenstates  and eigenvalues $\xi$-dependent), a family of non-regular measurements, referred to as controlled energy measurements, is naturally available. We perform an analytic optimization of their performance, which enables us to compare the optimal controlled energy measurement with the optimal Braunstein-Caves measurement based on the symmetric logarithmic derivative. As the former may outperform the latter, the ultimate quantum bounds for general Hamiltonian parameters are different than those for phase (shift) parameters. We also discuss in detail a realistic implementation of controlled energy measurements based on the quantum phase estimation algorithm and work out a variety of examples to illustrate our results. 
\end{abstract}  
\maketitle 
\section{Introduction}\label{s:intro} 
Quantum parameter estimation studies the statistical inference of an unknown parameter from the empirical data generated by a quantum system. The possible 
states of the system are described by a statistical model, i.e.~a family of 
density operators $\rho_\xi$, parametrized by $\xi$. An estimate of $\xi$ is obtained by performing a measurement on the system and then processing the 
outcomes via a point estimator $\hat \xi$ \cite{casella2002statistical,van2013detection,kay1993fundamentals,lehmann2006theory}. The overall task of parameter estimation is to optimize over the choice of both the measurement and the estimator, in order to minimize, on average, a given loss function. 
\par
The parameter to be estimated usually corresponds to a physical quantity which is not directly measurable. Quantum estimation is therefore  particularly relevant to the field of quantum technologies, since knowledge of inaccessible parameters is often required for quantum control. Following the pioneering works by Helstrom \cite{helstrom1976quantum} and Holevo \cite{holevo2011probabilistic}, it was discovered that estimation strategies exploiting quantum effects (such as 
squeezing \cite{caves1981quantum} and entanglement \cite{lop01,giovannetti2006quantum, giovannetti2011advances}) can outperform any classical strategy using the same resources (at least under ideal conditions \cite{demkowicz2012elusive,escher2011general,tsang2013quantum}). Quantum parameter estimation has thus become the theoretical foundation of quantum metrology \cite{akf95,giovannetti2004quantum,toth2014quantum,paris2009quantum}, besides being linked to branches of pure mathematics,  from statistics to information geometry \cite{hayashi2005asymptotic,amari2007methods,barndorff2003quantum}.   
\par
An important class of estimation problems is concerned with parameters characterizing the Hamiltonian $H_\xi$ of a closed quantum system. These
problems are referred to as Hamiltonian estimation problems, and the 
corresponding parameters $\xi$ as Hamiltonian parameters. One further 
distinguishes between phase (or shift) parameters and general 
parameters. In the former case, the parameter $\xi$ appears as an overall multiplicative constant, i.e.~$H_\xi= \xi G$, with $G$ being the generator 
of the system's dynamics. The phase estimation problem is well studied, 
both in the decoherence-free and noisy scenarios, with applications to 
optical interferometry, imaging and atomic spectroscopy \cite{huelga1997improvement,bollinger1996optimal,dowling1998correlated,dorner2009optimal,anisimov2010quantum,bri10,joo2011quantum,genoni2011optical,yonezawa2012quantum,kok2004quantum,pezze2009entanglement}. The case of a general Hamiltonian parameter, i.e.~when both the eigenvalues and the eigenvectors of $H_\xi$ depend on $\xi$, has been investigated only more recently \cite{brody2013information,pang2014quantum,seveso2017quantum,fraisse2017enhancing}.  
\par
In a Hamiltonian estimation problem, the system is initialized in the state 
$\rho_0$, the parameter is encoded through the unitary channel generated by $H_\xi$ and, finally, a measurement $\mathscr M$ is implemented. The outcomes of $N$ independent repetitions of the protocol are fed into an estimator $\hat \xi$, yielding an estimate of the parameter. If the estimator is unbiased and the loss function is the estimator's variance, then the performance of any estimation strategy is limited by the Cram\'er-Rao bound $$\text{Var}(\hat \xi)\geq [N \mathcal F_\xi(\rho_0,\,\mathscr M)]^{-1}\;,$$ where $\mathcal F_\xi$ denotes the Fisher information (FI) \cite{cramer2016mathematical,rao1992information,fisher1925theory}. The maximum of the FI over all possible initial preparations $\rho_0$ and measurements $\mathscr M$ is, by definition, the channel quantum Fisher information (CQFI) \cite{braunstein1994statistical}. The corresponding quantum Cram\'er-Rao bound $$\text{Var}(\hat \xi)\geq [N\mathcal F_\xi^{(Q,C)}]^{-1}$$ may be saturated by preparing the system in the optimal initial state, implementing the optimal measurement and processing the outcomes via an efficient estimator \footnote{Some caveats are necessary in order to clarify the previous proposition. If the optimal measurement depends on the true value of the parameter, an adaptive procedure is needed \cite{nagaoka1988asymptotically}. Moreover, if no unbiased efficient estimator exists, the Cram\'er-Rao bound can be saturated only in the asymptotic limit $N\to \infty$, by resorting to an asymptotically efficient estimator \cite{fujiwara2006strong}.}. The quantum Cram\'er-Rao is usually regarded as the ultimate quantum limit to precision, at least in the single parameter scenario \cite{barndorff2000fisher}, to which we will restrict our attention. 
\par
As argued in Ref.~\cite{seveso2017quantum}, in the estimation of a general Hamiltonian parameter, an enhanced precision limit is achievable. In a nutshell, the argument is as follows. The CQFI is the maximum FI, optimized over all initial preparations $\rho_0$ and over all \emph{regular} measurements, i.e.~measurements that are independent of the (unknown) true value of the parameter. The requirement that the measurement is parameter-independent is a natural assumption, analogous to the condition, for a classical statistical model $p_\xi$, that the support $\mathsf{supp} (p_\xi)$ is independent of $\xi$, which is a fundamental prerequisite for the Cram\'er-Rao bound to hold. Nonetheless, non-regular classical models have also been considered in the literature \cite{smith1985maximum,morimoto1967sufficient,akahira1975asymptotic1,akahira1975asymptotic2,woodroofe1972maximum,polfeldt1970order}. In such cases, an estimator performing better than predicted by the Cram\'er-Rao bound may exist \cite{akahira2012non}. Likewise, a quantum estimation strategy is referred to as regular if both the sample space $\mathcal X$ of possible outcomes and the POVM elements $\{\Pi_x\}_{x\in\mathcal X}$ are parameter-independent. In the context of a general Hamiltonian estimation problem, an energy measurement (i.e.~a projective measurement onto the eigenstates of the Hamiltonian $H_\xi$) is non-regular. In fact, both the outcomes of the measurement (the eigenvalues of $H_\xi$) and the detection operators (the projectors over its eigenstates) depend on $\xi$. The assessment of the best achievable precision becomes highly non-trivial in this case. In particular, the ultimate bound is no longer given by the CQFI. 
\par
Here, we further advance the analysis carried out in Ref.~\cite{seveso2017quantum}, specializing it to a class of 
non-regular measurements (referred to as {\em controlled energy measurements}) that arise in the estimation of a general Hamiltonian 
parameter. Having circumscribed the set of non-regular strategies under considerations, one is confronted with task of maximizing the precision over such a set. This is analogous to the introduction of the quantum Fisher information by a process of optimization over the set of regular measurements. One of the main results of the present manuscript is an analytic bound (which can be saturated under suitable conditions) to the best precision achievable via controlled energy measurements. We also discuss their experimental feasibility and propose a realistic implementation based on the quantum phase estimation algorithm. Finally, a collection of examples is employed to illustrate our results and emphasize that an enhancement (with respect to regular estimation strategies) can often be realized in practice.  
\par
The rest of the manuscript is organized as follows. Section \ref{s2} contains a basic review of quantum parameter estimation theory. In Section \ref{cemsec}, we introduce the family of controlled energy measurements, together with the information quantity $\mathcal G_\xi$, which quantifies the maximum extractable information in our setting. In Section \ref{bsec}, an upper-bound to $\mathcal G_\xi$ is derived, which is shown in Section \ref{ssec} to be tight for a large class of Hamiltonian problems. In Section \ref{exsec} we illustrate the relevance of our results to quantum metrology applications, showing how to perform a controlled energy measurement on a generic physical system. Finally, in Section 
\ref{S7}, a collection of examples is worked out. Section \ref{consec} 
closes the paper with some concluding remarks.
\section{Preliminaries}\label{s2}
We restrict ourselves to the case of a finite-dimensional quantum system with Hilbert space $\mathcal H = \mathbb C^d$. The generator of the system's noiseless evolution is its Hamiltonian $H_\xi\in \mathsf{Her}_d( \mathbb C)$, where $\mathsf{Her}_d( \mathbb C)$ is the set of $d\times d$ Hermitian matrices. The Hamiltonian $H_\xi$ depends generically on a parameter $\xi$, taking values in a parameter space $\Xi \subset \mathbb R$. Given a matrix $M \in \mathsf{Her}_d( \mathbb C)$, the following standard notation  is employed: $M$ has $d$ real eigenvalues $\mathsf{spec}(M)=\{\lambda_1(M),\,\dots,\,\lambda_d(M)\}$, ordered decreasingly, i.e.~$\lambda_1(M)\geq\dots\geq \lambda_d(M)$. The spectral gap $\sigma(M)$ is defined as the difference between its extremal eigenvalues, i.e.~$\sigma(M) \coloneqq \lambda_d(M)- \lambda_1(M)$. 

The computational basis of $\mathcal H$ is denoted by $\ket j$, with $j\in \{0,\mydots,d-1\}$, while the basis made up of the eigenstates of the Hamiltonian is denoted by $\ket{E_{j,\,\xi}}$; the subscript emphasizes that the energy eigenstates are $\xi$-dependent. By definition, $H_\xi\ket{E_{j,\,\xi}} = E_{j,\,\xi} \ket{E_{j,\,\xi}}$, where $E_{j,\,\xi}\coloneqq \lambda_{d-j}(H)$ are the eigenvalues of $H_\xi$. For simplicity, the spectrum of $H_\xi$ is assumed to be non-degenerate; however, everything that follows holds more generally also in the presence of degeneracies, with minor adaptations. For future convenience, we denote the projectors onto the computational basis (\emph{resp}., the energy eigenbasis)  by $P_{j} \coloneqq \ket{j}\bra{j}$ (\emph{resp}., $P_{E_{j,\,\xi}}\coloneqq \ket{E_{j,\,\xi}} \bra{E_{j,\,\xi}}$). The two basis are mapped one into the other by a suitable unitary similarity transformation $S_\xi$, i.e.~$\ket{j} = S_\xi \ket{E_{j,\,\xi}}$. Explicitly, the matrix elements of $S_\xi$ can be computed as $\braket{j|S_\xi|k} = \braket{E_{j,\,\xi}|k}$. Notice that $S_\xi$ reduces $H_\xi$ to diagonal form, i.e.~$S_\xi H_\xi S_\xi^\dagger = \mathsf{diag}(E_{0,\,\xi},\,\dots,\,E_{d-1,\,\xi})$, and that the matrix $S_\xi$ is $\xi$-dependent for a general Hamiltonian parameter. 

A typical quantum estimation strategy consists of the following steps: the system is initialized in the state $\rho_0$; then, the unitary map generated by $H_\xi$ encodes the parameter into the model $\rho_\xi \coloneqq U_t \rho_0 U_t^\dagger$, with $U_t \coloneqq \text{exp}(-itH_\xi)$ and $t$ the interrogation time; finally, a measurement $\mathscr M$ is performed. A measurement is defined in terms of its positive-operator valued measure (POVM) $\{\Pi_x\}_{x\in \mathcal X}$. Each $\Pi_x$ is a positive Hermitian operator, satisfying the completeness property $\sum_{x\in\mathcal X} \Pi_x = \mathbb I_d$, where $\mathbb I_d$ is the $d\times d$ identity matrix and the sample space $\mathcal X\subset \mathbb R$ is assumed to be a finite set. If both the sample space $\mathcal X$ and the POVM elements $\Pi_x$ do not depend on $\xi$, then the estimation strategy, as well as the measurement $\mathscr M$, are called \emph{regular}; the family of all possible regular measurements is denoted by $\rsfscr R$. Any given outcome $x\in\mathcal X$ is obtained with corresponding probability $p_{x,\,\xi} =\tr[\rho_\xi \Pi_x]$. Over $N$ repetitions of the protocol, one obtains a sample $\mathbf x \in \mathcal X^{\times N}$, which is processed via an estimator $\hat \xi\!: \mathcal X^{\times N}\to \xi$, yielding an estimate $\hat \xi(\mathbf x)$ of the parameter.  

Consider the set of all possible estimation strategies, with $\rho_0$ and $\mathscr M$ fixed, $\mathscr M \in \rsfscr R$ and $\hat \xi$ an unbiased estimator, i.e.~
\be
\mathbb E(\hat \xi) \coloneqq \sum_{\mathbf x\in \mathcal X^{\times N}} p_{\mathbf x,\,\xi}\, \hat \xi(\mathbf x)=\xi\;, \qquad \forall \xi \in \Xi\,.
\ee
where $p_{\mathbf x,\,\xi}$ is the joint probability distribution of the $N$ outcomes. Then, if the variance $\text{Var}(\hat \xi)$ is taken as the loss function, the Cram\'er-Rao theorem \cite{cramer2016mathematical,rao1992information,fisher1925theory} states that the best performing strategy, optimized over the choice of the estimator, saturates the inequality 
\be\label{crb}
\text{Var}(\hat \xi) \geq \frac{1}{N \cdot \mathcal F_{\xi}(\rho_0,\, \mathscr M)}\;.
\ee
The FI $\mathcal F_\xi(\rho_0,\,\mathscr M)$ is defined as follows
\be
\begin{split}
\mathcal F_\xi(\rho_0,\,\mathscr M) \coloneqq & \sum_{x\in\mathcal X} p_{x,\,\xi}\left(\partial_\xi \ln p_{x,\,\xi}\right)^2\\
=&\sum_{x\in\mathcal X} \tr[\rho_\xi \Pi_x]\,(\partial_\xi \ln\tr[\rho_\xi \Pi_x])^2\;.
\end{split}
\ee
We refer the interested reader to \cite{casella2002statistical,kay1993fundamentals,van2013detection} for precise conditions under which \eqref{crb} holds and can be achieved.

The next step is to optimize over the choice of the measurement $\mathscr M$. One introduces the QFI $\mathcal F_\xi^{(Q)}(\rho_0)$ as 
\be\label{qfidef}
\mathcal F_\xi^{(Q)}(\rho_0) = \underset{\mathscr M \in \rsfscr R}{\text{max}}\;\mathcal F_\xi(\rho_0,\,\mathscr M)\;.
\ee
Braunstein and Caves \cite{braunstein1994statistical,braunstein1995generalized} have proved that the QFI coincides with the least monotone quantum Riemannian metric in the Petz classification \cite{petz1996monotone}, which is the one based on the symmetric logarithmic derivative (SLD) \cite{helstrom1968minimum}, so that
\be
\mathcal F_\xi^{(Q)}(\rho_0) =\tr[\rho_\xi\, L_{\rho,\,\xi}^2]\;,\qquad \partial_\xi \rho_\xi = \frac{1}{2}\{\rho_\xi, L_{\rho,\,\xi}\}\;,
\ee
where $L_{\rho,\,\xi}$ is the SLD of $\rho_\xi$. Therefore, the best performing  strategy, optimized over the set of all (regular) measurements and unbiased estimators, for fixed initial preparation, saturates the inequality
\be
\text{Var}(\hat \xi ) \geq \frac{1}{N\cdot \mathcal F_\xi^{(Q)}(\rho_0)}\;.
\ee
Implementing the optimal Braunstein-Caves strategy requires performing a projective measurement over the eigenstates of $L_{\rho,\,\xi}$ and post-processing the outcomes via an efficient estimator \cite{fujiwara2006strong}. For pure models, i.e.~$\rho_0= \ket{\psi_0}\bra{\psi_0}$ and thus $\rho_\xi = \ket{\psi_\xi}\bra{\psi_\xi}$ with $\ket{\psi_\xi} = U_t \ket{\psi_0}$, the QFI can be computed explicitly \cite{paris2009quantum} as 
\be\label{qfivar}
\begin{split}
\mathcal F_\xi^{(Q)}(\rho_0) = &\, 4\, \text{Var}_{\ket{\psi_\xi}}\,\mathfrak g_{U,\,\xi}\\
=&\, 4\left[\braket{\psi_\xi|\mathfrak g_{U,\,\xi}^2|\psi_\xi}-\braket{\psi_\xi|\mathfrak g_{U,\,\xi}|\psi_\xi}^2\right]\;,
\end{split}
\ee 
where 
\be
\mathfrak g_{U,\,\xi} \coloneqq i\partial_\xi U_t U_t^\dagger
\ee
is the local generator of $U_t$ with respect to the parameter $\xi$. 

The final step is to optimize over the initial preparation $\rho_0$. The channel quantum Fisher information (CQFI) is defined as 
\be\label{cqfi}
\mathcal F_\xi^{(Q,\,C)}=\underset{\rho_0}{\text{max}}\, \mathcal F_\xi^{(Q)}(\rho_0)\;.
\ee
The QFI is a convex functional of the initial preparation, so that the maximum of Eq.~\eqref{cqfi} can be looked for on the set of pure states $\rho_0 = \ket{\psi_0}\bra{\psi_0}$. Since 
\be
\mathcal F_\xi^{(Q)}(\ket{\psi_0}\bra{\psi_0})= 4\, \text{Var}_{\ket{\psi_0}}\, U_t^\dagger\, \mathfrak g_{U,\,\xi}\, U_t\;,
\ee
 and moreover, by Popoviciu's inequality \cite{popoviciu1935equations}, the variance of a random variable $X$, with maximum value $x_M$ and minimum value $x_m$, is upper-bounded by $(x_M-x_m)^2/4$, it follows that
 \be 
 \begin{split}
\mathcal F_\xi^{(Q,\,C)}\leq & \, [\lambda_1(U_t^\dagger\, \mathfrak g_{U,\,\xi}\, U_t)-\lambda_d(U_t^\dagger\, \mathfrak g_{U,\,\xi}\, U_t)]^2\\
 = &\, [\lambda_1(\mathfrak g_{U,\,\xi})-\lambda_d(\mathfrak g_{U,\,\xi})]^2\;.
\end{split} 
\ee
Since a balanced superposition of the extremal eigenvectors of $\mathfrak g_{U,\,\xi}$ achieves the RHS of the previous inequality, the  inequality is tight and thus the CQFI is related to the spectral gap of the local generator $g_{U,\,\xi}$ via
 \be\label{cqfif}
\mathcal F_\xi^{(Q,\,C)} = (\sigma[\mathfrak g_{U,\,\xi}])^2\;.
 \ee
 Eq.~\eqref{cqfif} is the maximum information which can be extracted on $\xi$ via any regular quantum estimation strategy.  

\section{Non-regular estimation of Hamiltonian parameters}\label{cemsec}
A non-regular measurement depends intrinsically on the true value of the parameter, either via its sample space $\mathcal X_\xi$, or its POVM elements $\Pi_{x,\,\xi}$ (or both). The latter circumstance is specific to {quantum} parameter estimation, whereas the former has a classical analogue when the support of the statistical model $p_\xi$ is parameter-dependent. In such cases, it often happens that there exists a locally unbiased estimator with vanishing variance \cite{akahira2012non}, so that the achievable precision is formally unbounded. 
\par
As argued before, in the estimation of a general Hamiltonian parameter, a projective measurement of $H_\xi$ is non-regular, since either the eigenvalues, or the eigenstates of $H_\xi$, or both, depend on $\xi$. The first scenario would lead to a parameter-dependent sample space, the same situation one encounters in non-regular classical estimation. Assessing the performance of different strategies becomes a difficult matter; moreover, there is often no non-trivial lower bound to the variance of an unbiased estimator. In the rest of the manuscript, we will thus focus exclusively on the second case. That is, we are going to assume that either only the eigenvectors of $H_\xi$ depend on $\xi$, or that a data post-processing takes place after the energy measurement, which maps the original, parameter-dependent sample-space to a fixed, parameter-independent one. An energy measurement is thus modified by introducing a 1-1 map $\pi:\mathcal X_\xi\to \mathcal Y$, with $\mathcal X_\xi$ consisting of the eigenvalues of $H_\xi$, so that the sample space of the measurement is $\pi(\mathcal X_\xi)$, while its POVM elements are unchanged. The estimation strategy is still non-regular, but the possibility of pathological estimators with vanishing variance is excluded and the FI is again the relevant performance metric. 
\par
We now introduce a family of non-regular measurements $\mathscr M_{V,\,\xi}$, which is denoted by $\rsfscr E$; each measurement in $\rsfscr E$ is labelled by an arbitrary unitary control $V$. By definition, the measurement  $\mathscr M_{V,\,\xi}$ has POVM elements $V^\dagger P_{E_{j,\,\xi}}V$. It will be called a controlled energy measurement, since its implementation requires to apply a unitary, parameter-independent control $V$ to the system and thereafter measure its energy. 

In the absence of controls, a bare energy measurement ($V=\mathbb I_d$) obeys the statistics
\be
p_{E_{j,\,\xi}} = \tr[\rho_\xi P_{E_{j,\,\xi}}] = \braket{E_{j,\,\xi}|\rho_0|E_{j,\,\xi}}\;,
\ee
which does not depend on the interrogation time $t$. As a consequence, the corresponding FI $\mathcal F_\xi(\rho_0,\,\mathscr M_{\mathbb I_d,\,\xi})$ is also independent of $t$. In contrast, the QFI $\mathcal F_{\xi}^{(Q,\,C)}(\rho_0)$ grows generically like $t^2$ \cite{pang2014quantum}. If, however, a control is applied before the measurement, then the FI $\mathcal F_\xi(\rho_0,\, \mathscr M_{V,\,\xi})$ is again allowed to grow quadratically with $t$. This argument shows the metrological usefulness of controls in conjunction with an energy measurement.  

Finally, in analogy with the CQFI, we define the following information quantity
\be\label{cefid}
\mathcal G_\xi = \underset{\rho_0}{\text{max}}\;\underset{\mathscr M_{V,\,\xi}\in \rsfscr E}{\text{max}}\;\, \mathcal F_{\xi}(\rho_0,\,\mathscr M_{V,\,\xi})\;.
\ee
It represents the maximum extractable information on a general Hamiltonian parameter via controlled energy measurements, optimized over the set of initial preparations $\rho_0$ and unitary controls.
\begin{figure}[h]
\includegraphics[width=0.8\columnwidth]{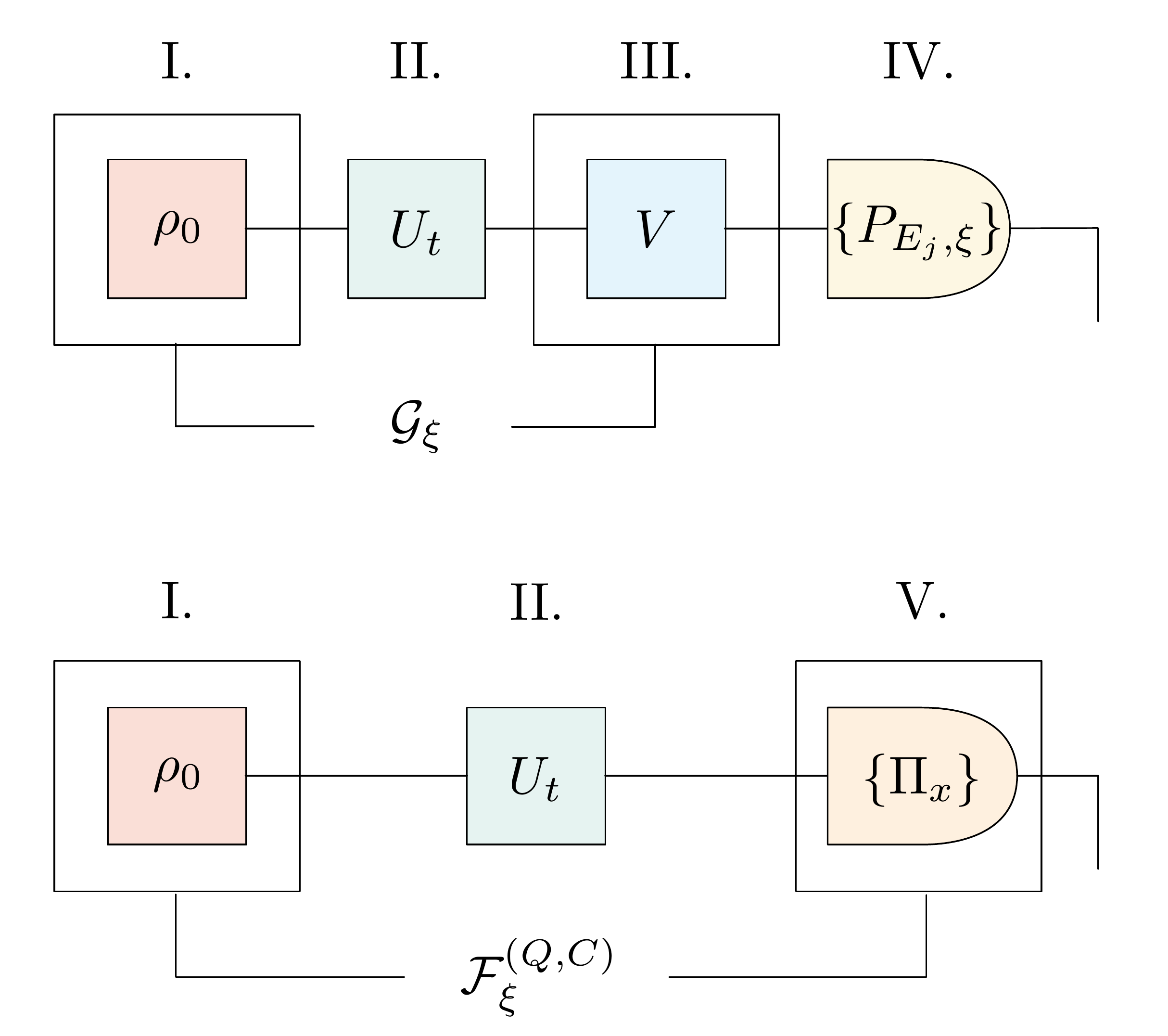}
\caption{Comparison between an estimation strategy based on a controlled energy measurement  (\emph{upper scheme}) and one based on a regular measurement (\emph{lower scheme}).
In the first case, the optimal performance is quantified by $\mathcal G_\xi$, 
which is optimized over the preparation and control steps; in the second, 
it is quantified by the CQFI $\mathcal F_\xi^{(Q,C)}$, which instead is 
optimized over all initial preparations and (regular) measurements. 
The different stages of the schemes are denoted as follows: I. $\rightarrow$ preparation, II. $\rightarrow$ encoding, 
III. $\rightarrow$ control, IV. $\rightarrow$ energy measurement and V. $\rightarrow$ regular measurement.
\label{f1s}}
\end{figure}\par
We summarize the preceding discussion via the following two definitions (see also Fig. \ref{f1s}):
\begin{defin}
Given a quantum system with Hamiltonian $H_\xi$ and unknown parameter $\xi\in \Xi$, a controlled energy measurement, denoted by $\mathscr M_{V,\,\xi}$, is defined through its POVM elements $ V^\dagger P_{E_{j,\,\xi}} V$, where $V$ is a unitary control and $P_{E_{j,\,\xi}}$ is the projector over the $j^{\text{th}}$ energy eigenstate. 
\end{defin}

\begin{defin}\label{def2}
The information quantity $\mathcal G_{\xi}$ is the maximum Fisher information $\mathcal F_{\xi}(\rho_0,\,\mathscr M_{V,\,\xi})$, optimized over both the set of initial preparations $\rho_0$ and controlled energy measurements $\mathscr M_{V,\,\xi}$.
\end{defin}

Let us remark that the performance of an estimation strategy making use of a controlled energy measurement is not necessarily bounded by the CQFI of Eq.~\eqref{cqfif}, i.e.~$\mathcal G_\xi$ may exceed $\mathcal F_\xi^{(Q,\,C)}$. However, computing $\mathcal G_\xi$ directly from its definition \eqref{cefid} is non-trivial. In the following, a closed-form expression for $\mathcal G_\xi$ (similar to Eq.~\eqref{cqfif} for $\mathcal F_\xi^{(Q,\,C)}$) is derived under the hypothesis that the Hamiltonian $H_\xi$ satisfies a rather general mathematical condition. For Hamiltonians not satisfying  such condition, it only provides an upper-bound to $\mathcal G_\xi$, which is not necessarily tight. With the help of this result, we will be able to compare regular estimation strategies with non-regular ones based on controlled energy measurements.

\section{Bounding $\mathcal G_\xi$}\label{bsec}
Consider a non-regular estimation strategy based on the controlled energy measurement $\mathscr M_{V,\,\xi}$. The probability distribution of the measurement outcomes is given by 
\be\label{sm}
\begin{split}
p_{\pi(E_{j,\,\xi}),\,\xi} = & \tr\left[\rho_\xi\, V^\dagger P_{E_{j,\,\xi}} V\right] \\
= & \tr\left[(S_\xi V U_t)\, \rho_0\, (S_\xi V U_t)^\dagger  P_{j} \right] \\
= & \tr\left[\mathcal U_{V} \rho_0\, \mathcal U_{V}^\dagger P_{j} \right]\;,
\end{split}
\ee
where all dependence on $\xi$ has been collected in the unitary matrix $\mathcal U_{V} \coloneqq S_\xi V U_t$. We define the statistical model $\rho_{V,\,\xi}\coloneqq \mathcal U_{V} \rho_0\, \mathcal U_{V}^\dagger$ as the model which one would obtain if the parameter were encoded on the initial preparation $\rho_0$ through $\mathcal U_{V}$, instead of $U_t$; it is referred to as the {auxiliary statistical model} associated to the physical model $\rho_\xi$. It follows from Eq.~\eqref{sm} that the FI $\mathcal F_\xi (\rho_0,\,\mathscr M_{V,\,\xi})$, for the non-regular estimation strategy we are considering, is formally equal to the FI corresponding to a projective measurement in the computational basis on the auxiliary model, i.e.
\be\label{fiam}
\mathcal F_\xi (\rho_0,\,\mathscr M_{V,\,\xi}) = \sum_{j=0}^{d-1} \frac{\left(\partial_\xi \tr[\rho_{V,\,\xi} P_{j}]\right)^2}{\tr[\rho_{V,\,\xi} P_{j}]}\;.
\ee 
Following Braunstein and Caves \cite{braunstein1994statistical}, the Fisher information \eqref{fiam} can be majorized as follows. After expressing the derivative at the numerator as $\partial_\xi \rho_{V,\,\xi} = \{\rho_{V,\,\xi},\, L_{\rho_V,\,\xi}\}/2$, where $L_{\rho_V,\,\xi}$ is the SLD of the auxiliary model, one obtains
\be\label{bstuff}
\begin{split}
\mathcal F_\xi(\rho_0,\,\mathscr M_{V,\,\xi}) & = \frac{1}{2}\sum_{j=0}^{d-1} \frac{\left(\tr\left[\{\rho_{V,\,\xi}, L_{\rho_V,\,\xi}\}\,P_j\right]\right)^2}{\tr\left[\rho_{V,\,\xi}\,P_j\right]}\\
&= \sum_{j=0}^{d-1} \frac{\Re^2\left(\tr\left[ \rho_{V,\,\xi}\; L_{\rho_V,\,\xi}\,P_j\right]\right)}{\tr\left[\rho_{V,\,\xi}\,P_j \right]}\\
&\leq \sum_{j=0}^{d-1} \frac{\left|\tr\left[\rho_{V,\,\xi}\; L_{\rho_V,\,\xi}\,P_j\right]\right|^2}{\tr\left[\rho_{V,\,\xi}\, P_j\right]}\;,
\end{split}
\ee
where use was made of the triangular inequality $\Re z\leq \abs{z},\;\forall z\in\mathbb C$. 

Next, by the Cauchy-Schwarz inequality, the numerator can be bounded as follows, 
\be\label{csin}
\abs{\tr\left[\rho_{V,\,\xi}\, L_{\rho_V,\,\xi}P_j\right]}^2 
\leq \tr\left[L_{\rho_V,\,\xi}\,\rho_{V,\,\xi}\, L_{\rho_V,\,\xi} P_j\right] \tr\left[\rho_{V,\,\xi}\, P_j\right]\;.
\ee
Therefore, 
\be\label{bb}
\begin{split}
\mathcal F_\xi(\rho_0,\,\mathscr M_{V,\,\xi}) \leq \, & \sum_{j=0}^{d-1} \tr\left[L_{\rho_V,\,\xi}\,\rho_{V,\,\xi}\, L_{\rho_V,\,\xi} P_j\right] \\ = \, & \tr\left[\rho_{V,\,\xi}\,  (L_{\rho_V,\,\xi})^2\right]\;.
\end{split}
\ee
Taking the maximum over the initial preparation, 
\be
\underset{\rho_0}{\text{max}}\;\mathcal F_\xi(\rho_0,\,\mathscr M_{V,\,\xi}) \leq \underset{\rho_0}{\text{max}}\; \tr\left[\rho_{V,\,\xi}\,  (L_{\rho_V,\,\xi})^2\right]\;.
\ee
By convexity, the maximum of the expression on the RHS is achieved for a pure initial preparation, i.e.~$\rho_0 = \ket{\psi_0}\bra{\psi_0}$. On the other hand, for pure initial preparation, one can rewrite it as
\be
\tr\left[\rho_{V,\,\xi}\,  (L_{\rho_V,\,\xi})^2\right]_{ \rho_0=\ket{\psi_0}\bra{\psi_0}} = 4\text{Var}_{\ket{\psi_0}}(\mathcal U_{V}^\dagger\, \mathfrak g_{\mathcal U_V,\,\xi}\, \mathcal U_{V})\;,
\ee
where 
\be\label{deco}
\mathfrak g_{\mathcal U_V,\,\xi} = \mathfrak g_{S,\,\xi} + (S_{\xi} V)\,\mathfrak g_{U,\,\xi}\,(S_{\xi} V)^\dagger
\ee
is the local generator of the unitary encoding $\mathcal U_{V}$ for the auxiliary model. By Popoviciu's inequality, it follows that 
\be\label{s2in}
\underset{\rho_0}{\text{max}}\;\mathcal F_\xi(\rho_0,\,\mathscr M_{V,\,\xi}) \leq \left(\sigma\left[ \mathfrak g_{S,\,\xi} + (S_{\xi} V)\,\mathfrak g_{U,\,\xi}\,(S_{\xi} V)^\dagger\right]\right)^2\;.
\ee

Finally, one maximizes over the choice of the unitary control $V$, i.e.
\be\label{rhs1}
\mathcal G_\xi \leq \underset{V\in U(d)}{\text{max}} \left(\sigma\left[ \mathfrak g_{S,\,\xi} + (S_{\xi} V)\,\mathfrak g_{U,\,\xi}\,(S_{\xi} V)^\dagger\right]\right)^2\;.
\ee
The maximization  on the RHS can be carried out explicitly with the help of the following lemma.

\begin{lem}\label{ml}
Given two Hermitian matrices $M_1,\,M_2 \in \mathsf{Her}_d( \mathbb C)$, the maximum spectral gap of the sum of any other two Hermitian matrices $\tilde M_1,\,\tilde M_2$ with the same spectra (i.e.~$\text{spec}(M_i)=\text{spec}(\tilde M_i)$, $i=1,2$) is equal to the sum of the spectral gaps $\sigma(M_1)+\sigma(M_2)$:
\be
\underset{\substack{\tilde M_1,\,\tilde M_2}}{\text{\emph{max}}}\,\sigma(\tilde M_1 + \tilde M_2) = \sigma(M_1)+\sigma(M_2)\;.
\ee

\begin{proof}
One may write $\tilde M_i = U_i M_i U_i^\dagger$, for suitable unitary matrices $U_i$, ($i=1,2$). Since the spectral gap is invariant under unitary transformations, it follows that
\be
\sigma(U_1 M_1 U_1^\dagger + U_2 M_2 U_2^\dagger) = \sigma(M_1 + U M_2 U^\dagger)\;,
\ee
where $U\coloneqq  U_1^\dagger U_2$. Therefore, we have to prove that
\be
\underset{U\in U(d)}{\text{max}}\,\sigma(M_1 + U M_2 U^\dagger) = \sigma(M_1)+\sigma(M_2)\;,
\ee
By definition, the LHS is equal to
\be
\underset{U\in U(d)}{\text{{max}}}\; [\lambda_1(M_1+U M_2 U^\dagger)-\lambda_d(M_1+U M_2 U^\dagger)]\;.
\ee
The first term may be bounded as follows, 
\be
\begin{split}
\lambda_1(M_1+U M_2 U^\dagger)=&\, \underset{\ket \psi}{\text{max}}\, \braket{\psi|M_1+U M_2 U^\dagger|\psi}\\
\leq &\, \underset{\ket \psi}{\text{max}}\, \braket{\psi|M_1|\psi}+\underset{\ket \psi}{\text{max}}\,\braket{\psi|U M_2 U^\dagger|\psi}\\
= &\, \underset{\ket \psi}{\text{max}}\, \braket{\psi|M_1|\psi}+\underset{\ket \psi}{\text{max}}\,\braket{\psi| M_2|\psi}\\
=&\, \lambda_1(M_1)+\lambda_1(M_2)\;.
\end{split}
\ee
Similarly, one proves that
\be
\lambda_d(M_1+U M_2 U^\dagger)\geq \lambda_d(M_1)+\lambda_d(M_2)\;.
\ee
The last two inequalities imply that
\be\label{ltiit}
\sigma(M_1 + U M_2 U^\dagger) \leq \sigma(M_1)+\sigma(M_2)\;.
\ee
What is left to prove is that the bound is tight. Choose $U = R_1^\dagger\,R_2$, where $R_1$ (\emph{resp.}, $R_2$) is the similarity transformation which diagonalizes $M_1$ (\emph{resp.}, $M_2$), with the eigenvalues ordered decreasingly, i.e.
\be
\begin{split}
R_1 M_1 R_1^\dagger=\,&\mathsf{diag}(\lambda_1(M_1),\dots,\lambda_d(M_1))\coloneqq D_1\;,\\
R_2 M_2 R_2^\dagger=\,&\mathsf{diag}(\lambda_1(M_2),\dots,\lambda_d(M_2))\coloneqq D_2\;.
\end{split}
\ee
Then, for this particular choice of $U$,
\be
\begin{split}
\lambda_1(M_1+U M_2 U^\dagger) =& \lambda_1(R_1 M_1 R_1^\dagger + R_2 M_2 R_2^\dagger)\\ =& \lambda_1(D_1)+\lambda_1(D_2) \\ =& \lambda_1(M_1) + \lambda_1(M_2)\;.
\end{split}
\ee
 Similarly, one finds that
\be
\lambda_d(M_1+U M_2 U^\dagger) = \lambda_d(M_1) + \lambda_d(M_2)\;.
\ee
Therefore, the RHS of \eqref{ltiit} is achievable. 
\end{proof}
\end{lem}

Using the lemma, it follows that the RHS of Eq.~\eqref{rhs1} is equal to $\left(\sigma[\mathfrak g_{U,\,\xi}]+\sigma[\mathfrak g_{S,\,\xi}]\right)^2$. We have therefore established the following proposition:
\begin{prop}\label{pr1}
Given a finite-dimensional quantum system with Hamiltonian $H_\xi \in \mathsf{Her}_d( \mathbb C) $ and general parameter $\xi \in \Xi$, the performance of any non-regular  estimation strategy based on a controlled energy measurement $\mathscr M_{V,\,\xi}$ is bounded as follows. The maximum extractable information $\mathcal G_\xi$ obeys the inequality 
\be
\label{final}
\mathcal G_\xi \leq\left(\sigma[\mathfrak g_{U,\,\xi}]+\sigma[\mathfrak g_{S,\,\xi}]\right)^2\;,
\ee
where $U_t =  \text{exp}(-itH)$ is the unitary encoding, $S_{\xi}$ is the similarity transformation diagonalizing $H_\xi$, $\mathfrak g_{U,\,\xi}$ (\emph{resp.}, $\mathfrak g_{S,\,\xi}$) is the generator of $U_t$ (\emph{resp.}, $S_{\xi}$), i.e.
\be
\mathfrak g_{U,\,\xi} = i\partial_\xi U_t U_t^\dagger\;,\qquad\qquad \mathfrak g_{S,\,\xi} = i\partial_\xi S_{\xi} S_{\xi}^\dagger\;,
\ee
and $\sigma(\cdot)$ denotes the spectral gap. 
\end{prop}

\section{Saturating the inequality in Eq. (\ref{final})}
\label{ssec}
If the eigenvectors of $H_\xi$ do not actually depend on $\xi$ (so that $\partial_\xi S_\xi = 0$) then the set of controlled energy measurements coincides with that of (parameter-independent) projective measurements; since the CQFI is achieved for a projective measurement, it follows that $\mathcal G_\xi$ must reduce to the CQFI $\mathcal F_\xi^{(Q,\,C)}$. On the other hand, if $\partial_\xi S_\xi = 0$, then $\sigma(\mathfrak g_{S,\,\xi})=0$, so the RHS of inequality~\eqref{final} is also equal to the CQFI (by comparison with Eq.~\eqref{cqfi}). Thus, at least in such limiting case, the inequality $\mathcal G_\xi \leq \left(\sigma[\mathfrak g_{U,\,\xi}]+\sigma[\mathfrak g_{S,\,\xi}]\right)^2$ is saturated. In this section, we discuss under which general conditions the bound given in Prop.~\ref{pr1} can be tight. We discover that tightness depends only on a mathematical property of the Hamiltonian $H_\xi$, explained below. Therefore, for all Hamiltonians belonging to such special class, $\mathcal G_\xi$ can be readily computed in terms of the spectral gaps of the generators of $U_t$ and $S_\xi$. 

Let us summarize the steps that went into proving the bound of Prop.~\ref{pr1}. First, we bounded the FI from above, in Eq.~\eqref{bb} ($\mathsf{step \;1}$). Next, we optimized over the initial preparation, which led to Eq.~\eqref{s2in} ($\mathsf{step \;2}$). Finally, we optimized over the unitary control $V$ by means of Lemma \ref{ml} ($\mathsf{step \;3}$). The last two steps were proper maximizations, so they can be made tight by implementing the optimal control $V_{opt}$ and the optimal initial preparation $\ket{\psi_{0,\,opt}}$. The optimal control $V_{opt}$, which achieves the maximum in $\mathsf{step \;3}$, is obtained from the proof of Lemma \ref{ml}: 
\be\label{opt1}
V_{opt} = S_\xi^\dagger\, R_{1}^\dagger R_{2}\;,
\ee 
where $R_{1}$ (\emph{resp.}, $R_{2}$) is the similarity transformation which diagonalizes $\mathfrak g_{S,\,\xi}$ (\emph{resp.}, $\mathfrak g_{U,\,\xi}$), with eigenvalues ordered decreasingly, i.e.
\begin{equation*}
\begin{split}
R_{1} \,\mathfrak g_{S,\,\xi}\, R_{1}^\dagger =\,& \mathsf{diag}\left[\lambda_1(\mathfrak g_{S,\,\xi}),\dots,\lambda_d(\mathfrak g_{S,\,\xi})\right]\;,\\ 
R_{2} \,\mathfrak g_{U,\,\xi}\, R_{2}^\dagger =\,& \mathsf{diag}\left[\lambda_1(\mathfrak g_{U,\,\xi}),\dots,\lambda_d(\mathfrak g_{U,\,\xi})\right]\;.
\end{split}
\end{equation*}
The optimal preparation $\ket{\psi_0}_{opt}$, which achieves the maximum in $\mathsf{step \;2}$, is related to the extremal eigenvectors of $\mathcal U_{opt}^\dagger\, \mathfrak g_{\mathcal U_{opt},\,\xi}\, \mathcal U_{opt}$, where $\mathcal U_{opt} \coloneqq S_\xi V_{opt} U_t$ and $\mathfrak g_{\mathcal U_{opt},\,\xi}$ is its generator. Explicitly,
\be\label{opt2}
\ket{\psi_{0,\,opt}} = \frac{1}{\sqrt{2}}\,\mathcal U_{opt}^\dagger \left[\ket{\lambda_1(\mathfrak g_{\mathcal U_{opt},\,\xi})}+e^{i\varphi}\ket{\lambda_d(\mathfrak g_{\mathcal U_{opt},\,\xi})}\right]\;,
\ee 
where $\varphi \in \mathbb R$.

Tightness of inequality \eqref{final} is thus reduced to that of $\mathsf{step \;1}$, with the control and the initial preparation chosen according to Eqs.~\eqref{opt1} and \eqref{opt2}, respectively. In turn, $\mathsf{step \;1}$ involves two majorizations. The first majorization is based on the Cauchy-Schwarz inequality of Eq.~\eqref{csin}, which is saturated iff
\be
\label{cc}
 \sqrt{\rho_{opt}}\,P_j \propto \sqrt{\rho_{opt}}\, L_{\rho_{opt},\,\xi}\,P_j\;,\qquad\forall j\in \{0,\mydots,d-1\}\;,
\ee
where $\rho_{opt}\coloneqq \mathcal U_{opt}\,\rho_{0,\,opt}\,\mathcal U_{opt}^\dagger$, $L_{\rho_{opt},\,\xi}$ is its SLD and the proportionality constant may depend on $j$. Condition \eqref{cc} is always satisfied thanks to the fact that the model is pure, i.e.~$\ket{\psi_{opt}}\coloneqq \mathcal U_{opt} \ket{\psi_{0,\,opt}}$, since then it reduces to the manifestly true relation
\be
\braket{\psi_{opt}|j}\, \ket{\psi_{opt}}\bra{j}\propto \bra{\psi_{opt}}\,L_{\psi_{opt},\,\xi} \ket{j}\, \ket{\psi_{opt}}\bra{j}\;.
\ee 

The second majorization is based on the triangular inequality of Eq.~\eqref{bstuff}. It is proven below that saturation occurs iff the unitary matrix $S_\xi$ has \emph{equioriented} extremal eigenvectors (two complex vectors $v_1,\,v_2$ are said to be equioriented, with respect to a given orthonormal basis $\ket{b_j}$, if $\,\abs{\braket{b_j|v_1}}=\abs{\braket{b_j|v_2}}$, $\forall j \in \{0,\mydots,d-1\}$). For all Hamiltonians such that the matrix $S_\xi$ has the previous property, inequality \eqref{final} becomes an equality. We collect our results in the following two propositions.
\begin{prop}\label{pr2}
The inequality given in Prop.~\ref{pr1} is an equality when the Hamiltonian $H_\xi$ is such that the extremal eigenvectors of the generator $\mathfrak g_{S,\,\xi}$ of $S_{\xi}$ are equioriented with respect the computational basis. 
\begin{proof} 
Most of the proof is contained in the discussion preceding Eq.~\eqref{pr2}. What is left to check is that the triangular inequality of Eq.~\eqref{bstuff} is saturated whenever $\mathfrak g_{S,\,\xi}$ has equioriented extremal eigenvectors. For Eq.~\eqref{bstuff} to be tight, it must be that, $\forall j\in \{0,\mydots,d-1\}$, 
\be
\Im \tr\left(\rho_{opt}\, L_{\rho_{opt},\,\xi}\, P_j\right) = 0\;,
\ee
which is also equivalent to
\be\label{crpe0}
\Im \left[\bra{j}L_{\psi_{opt},\,\xi}\ket{\psi_{opt}}\braket{\psi_{opt}|j}\right] = 0\;.
\ee
The SLD $L_{\psi_{opt},\,\xi}$ is given by
\be
L_{\psi_{opt},\,\xi} = 2 \ket{\partial_\xi \psi_{opt}} \bra{\psi_{opt}} + 2 \ket{\psi_{opt}} \bra{\partial_\xi\psi_{opt}}\;,
\ee
which can be rewritten as
\be 
L_{\psi_{opt},\,\xi}\ket{ \psi_{opt}} = 2i\left(\braket{\psi_{opt}|\mathfrak g_{\mathcal U_{opt},\,\xi}|\psi_{opt}}-\mathfrak g_{\mathcal U_{opt},\,\xi}\right)\ket{\psi_{opt}}\;.
\ee
Substituting this result in Eq.~\eqref{crpe0}, one arrives at the condition
\be
\braket{\psi_{opt}|\mathfrak g_{\mathcal U_{opt},\,\xi}|\psi_{opt}}\abs{\braket{j|\psi_{opt}}}^2 \!= \Re[\braket{j|\mathfrak g_{\mathcal U_{opt},\,\xi}|\psi_{opt}}\!\braket{\psi_{opt}|j}]
\ee
or, using the explicit form of the optimal preparation given in Eq.~\eqref{opt2}, 
\be
\begin{split}
0= & \left(\abs{\braket{j|\lambda_1(\mathfrak g_{\mathcal U_{opt},\,\xi})}}^2 -\abs{\braket{j|\lambda_d(\mathfrak g_{\mathcal U_{opt},\,\xi})}}^2\right)\\
 &\qquad\times[\lambda_1(\mathfrak g_{\mathcal U_{opt},\,\xi})-\lambda_d(\mathfrak g_{\mathcal U_{opt},\,\xi})]\;.
\end{split}
\ee
The conclusion is that the extremal eigenvectors of $\mathfrak g_{\mathcal U_{opt},\,\xi}$ must be equioriented. To finish the proof, we have to show that $\ket{\lambda_i(\mathfrak g_{\mathcal U_{opt},\,\xi})} = \ket{\lambda_i(\mathfrak g_{S })}$ for $i=1,\,d$. This can be proven as follows. Note that
\be
\begin{split}
\mathfrak g_{\mathcal U_{opt},\,\xi} =\; &\mathfrak g_{S,\,\xi} + R_{1}^\dagger R_{2}\,\mathfrak g_{U,\,\xi}\, R_{2}^\dagger R_{1} = R_{1}^\dagger\,D\,R_{1} \;,
\end{split}
\ee 
where $D$ is the diagonal matrix 
\ben
D=\mathsf{diag}[\lambda_1(\mathfrak g_{S,\,\xi})+\lambda_1(\mathfrak g_{U,\,\xi}),\dots,\lambda_d(\mathfrak g_{S,\,\xi})+\lambda_d(\mathfrak g_{U,\,\xi})]\;.
\een
Therefore, the extremal eigenvectors of $\mathfrak g_{\mathcal U_{opt},\,\xi}$ are given by $R_{1}^\dagger\ket{1}$ and $R_{1}^\dagger\ket{d}$. But, by definition of $R_{1}$, these are also the extremal eigenvectors of $\mathfrak g_{S,\,\xi}$. 
\end{proof}
\end{prop}

\begin{prop}
If the conditions of Prop.~\ref{pr2} are satisfied, the strategy which saturates the bound \eqref{final} makes use of the optimal control $V_{opt} = S_\xi^\dagger\, R_{1}^\dagger R_{2} $ and the optimal initial preparation $\ket{\psi_{0,\,opt}}$, i.e.
\ben
\ket{\psi_{0,\,opt}} = \frac{1}{\sqrt{2}} (S_\xi V_{opt} U_t)^\dagger \left[\ket{\lambda_1(\mathfrak g_{S,\,\xi})}+e^{i\varphi}\ket{\lambda_d(\mathfrak g_{S,\,\xi})}\right]\;,
\een
where $\varphi \in \mathbb R$ and $R_{1}$ (\emph{resp.}, $R_{2}$) is the similarity transformation which diagonalizes $\mathfrak g_{S,\,\xi}$ (\emph{resp.}, $\mathfrak g_{U,\,\xi}$), with eigenvalues ordered decreasingly.
\begin{proof}
Follows immediately from Eq.~\eqref{opt1} and \eqref{opt2}, together with the fact that $\mathfrak g_{\mathcal U_{opt},\,\xi}$ and $\mathfrak g_{S,\,\xi}$ have the same extremal eigenvectors.
\end{proof}
\end{prop}

The condition that the extremal eigenvectors of $\mathfrak g_{S,\,\xi}$ be equioriented might appear restrictive, but actually it is often satisfied in practice (see also Sect.~\ref{S7}). In such cases, the LHS of Eq.~\eqref{final} provides a simple expression for $\mathcal G_\xi$. The possible precision enhancement with respect to the optimal Braunstein-Caves measurement is then quantified by
\be
\Delta = \left(\sigma[\mathfrak g_{U,\,\xi}]+\sigma[\mathfrak g_{S,\,\xi}]\right)^2-\sigma[\mathfrak g_{U,\,\xi}]^2\;.
\ee

\section{Application to metrology}\label{exsec}
In this section, we illustrate the relevance of our previous results to quantum metrology applications. The main point to address is how to perform a controlled energy measurement on a physical system. In principle, one is required to 
initialize the system in a reference state $\rho_0$, to encode the parameter $\xi$, to apply a unitary control $V$ and finally to measure the energy. The problem is that the Hamiltonian is not fully known and, as a result, neither is the POVM to 
be implemented.
\par
When the Hamiltonian is fully known, a projective energy measurement can be performed by a suitable modification of the phase estimation algorithm \cite{temme2011quantum,riera2012thermalization,kitaev1995quantum}. However, 
such an approach is not useful for parameter estimation, since it requires 
to know the parameter beforehand. Let us also emphasize that, for a controlled energy measurement to be non-regular, it is crucial that the measurement projects onto the eigenstates of the \emph{true} Hamiltonian $H_\xi$. Otherwise, the measurement is regular (and thus cannot outperform the optimal Braunstein-Caves measurement). In conclusion, our aim is to design a measurement such that its statistics coincides (or at least approximates closely) that of a controlled 
energy measurement \emph{for all} $\xi \in \Xi$ assuming no knowledge about the system's Hamiltonian.
\par
\begin{figure*}
\includegraphics[width=0.7\textwidth]{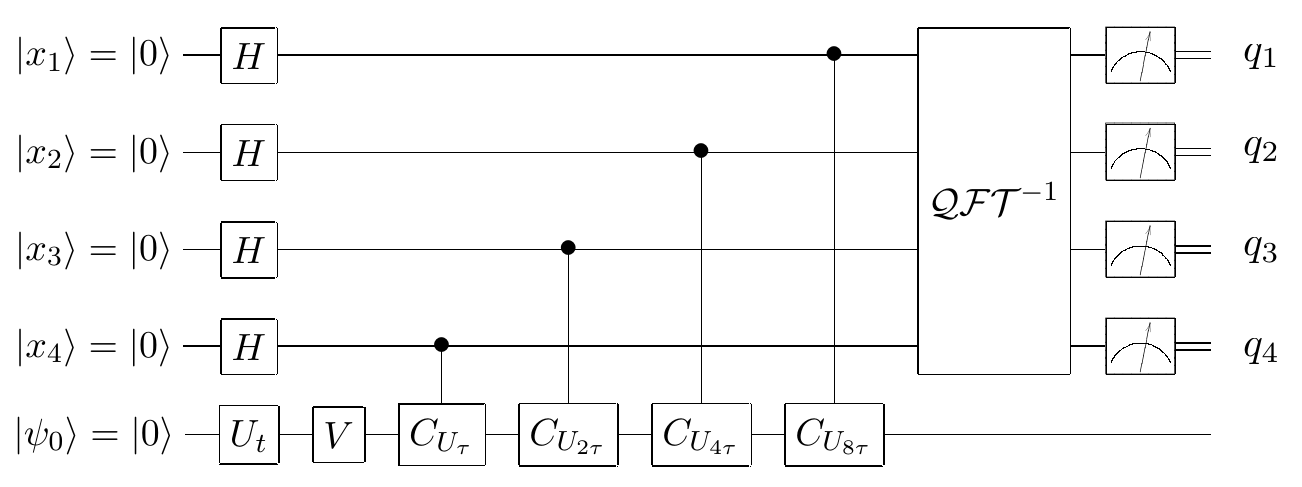}
\caption{Circuit diagram of a simplified realistic controlled energy measurement with $n=4$ control qubits. A realistic controlled energy measurement replaces each application of $\mathcal C_{U_\tau}$ by $m$ repeated applications of $\Gamma_{U_{\tau/m}}$,  defined in Eq.~\eqref{biggamma}. \label{algo}}
\end{figure*}

Let us now explain how to construct such a measurement, referred to as a realistic controlled energy measurement. The central idea is to make use of the system's unitary evolution as a resource, by means of a quantum subroutine, named {\em universal controllization} and developed in \cite{nakayama2015quantum,matsuzaki2017projective}. First, we describe a 
simplified version of a controlled energy measurement (see Fig. \ref{algo}), 
which is 
actually based on an unrealistic assumption; then, we explain how to 
remove such assumption. The assumption is that the experimenter can 
implement the controlled time-evolution operator
\be
C_{U_t} \eqd \ket{0}\bra{0} \otimes \mathbbm {I}_d + \ket{1}\bra{1} \otimes U_t\;,
\ee
acting on the enlarged Hilbert space $\mathbb C^2 \otimes \mathcal H$, where $\mathcal H = \mathbb C^d$ is the Hilbert space of the main system and $U_t = \text{exp}(-it H_\xi)$ is the time-evolution operator. The assumption is unrealistic because $C_{U_t}$ still depends on the true value of the parameter $\xi$, which is not known. 
\par
In order to implement this simplified version, let us introduce $n$ control qubits, each with Hilbert space $\mathcal H_c = \mathbb C^2$; the total Hilbert space is now $\mathcal H_c^{\otimes n} \otimes \mathcal H$. Each control qubit is prepared in the ground state $\ket{0}$. Thus, the initial state of the system is  $\ket{0\mydots 0} \bra{0\mydots 0}\otimes \rho_0$. Next, a Hadamard gate is applied to each control qubit, i.e. $\ket{0} \to H \ket{0} =(\ket{0} +\ket{1})/\sqrt 2$. Meanwhile, the parameter is encoded into $\rho_\xi = U_t \rho_0 U_t^\dagger$ and the unitary control $V$ is applied. Therefore, the state of the total system up to this step is
\be
\frac{1}{2^{n}}\,\sum_{x,y \in \{0,1\}^{\times n}}
\ket{x_1\mydots\, x_n} \bra{y_1\mydots\, y_n} \otimes V \rho_\xi V^\dagger\;,
\ee
where $x$ stands for the generic binary $n$-string $x_1\mydots\, x_n$. 
\par
Given any unitary $U$ acting on $\mathcal H$, the superoperator $\mathcal C_U$ is defined as follows, 
\be\label{CU}
\mathcal C_U [\rho] \eqd C_U \rho\, C_U^\dagger\;. 
\ee
For $l=1,\mydots,n$, the $n$ superoperators $\mathcal C_{U_{\tau}^{2^{l-1}}}$ are applied between the $l^{\text{th}}$ control qubit and the main system ($\tau$ is a free parameter giving the timescale of the measurement process). Notice that, when $\mathcal C_{U_{\tau}^{2^{l-1}}}$ is applied to $\rho_l \eqd \ket{x_l}\bra{y_l}\otimes V \rho_\xi V^\dagger$, one obtains
\be
\mathcal C_{U_{\tau}^{2^{l-1}}}\left[\rho_l \right] =  \ket{x_l} \bra{y_l} \otimes  U_\tau^{x_l 2^{l-1}} V \rho_\xi V^\dagger \left(U^\dagger_\tau\right)^{y_l 2^{l-1}}\;.
\ee
Denoting by $X = x_1 + 2\cdot x_2 +\mydots + 2^{n-1}\cdot x_n $ the decimal representation of the binary string $x$, the resulting total state is 
\be\label{ts1}
\frac{1}{2^{n}}\, \sum_{X=0}^{2^n-1}\, \sum_{Y=0}^{2^n-1} \ket{x} \bra{y} \otimes U_\tau^X V \rho_\xi V^\dagger (U^\dagger_\tau)^Y\;.
\ee
Let us expand $V\rho_\xi V^\dagger$ on the energy eigenbasis, i.e.
\be
V\rho_\xi V^\dagger = \sum_{j=0}^{d-1} \sum_{k=0}^{d-1} c_{jk} \ket{E_{j,\,\xi}} \bra{E_{k,\,\xi}}\;.
\ee
Eq.~\eqref{ts1} then becomes
\be
\frac{1}{2^{n}}\! \sum_{j,k=0}^{d-1}\, \sum_{X,Y=0}^{2^n-1} c_{jk}\,e^{-i\tau (X E_{j,\,\xi}-Y E_{k,\,\xi})}
\ket{x}\! \bra{y} \otimes \ket{E_{j,\,\xi}}\! \bra{E_{k,\,\xi}}.
\ee
The next step is to apply an inverse quantum Fourier transform $\mathcal{QFT}^{-1}$ on the $n$ control qubits. By definition, $\mathcal{QFT}^{-1}$ acts as follows on the computational basis of $\mathcal H_c^{\otimes n}$:
\be
\mathcal{QFT}^{-1}\ket{x} = \frac{1}{2^{n/2}}\, \sum_{Q=0}^{2^{n}-1} e^{-\frac{2\pi i X Q}{2^n}}\ket{q}\,.
\ee
After application of $\mathcal{QFT}^{-1}$, the total state of the system is
\be
\frac{1}{2^{2n}}\, \sum_{j,k=0}^{d-1}\,\sum_{X,Y=0}^{2^n-1}\,\sum_{Q,P=0}^{2^n-1}\, \tilde c_{jk}\, \ket{q}\bra{p} \otimes \ket{E_{j,\,\xi}}\bra{E_{k,\,\xi}}\;.
\ee
where 
\be
\tilde c_{jk} = c_{jk}\, e^{-iX\left(\tau E_{j,\,\xi}+\frac{2\pi Q}{2^n}\right)}\, e^{iY\left(\tau E_{k,\,\xi}+\frac{2\pi P}{2^n}\right)}\;.
\ee
The last step is to perform a measurement of the $n$ control qubits in the computational basis. The probability $p_{q,\,\xi}$ of obtaining as outcome the binary string $q$ is 
\be\label{pp2}
p_{q,\,\xi} = \frac{1}{2^{2n}} \sum_{j=0}^{d-1} \, \sum_{X,Y=0}^{2^n-1} p_{E_j,\,\xi}\, e^{-i(X-Y)\alpha_{j,Q}}\;,
\ee
where
\be
\alpha_{j,Q} \eqd \tau E_{j,\,\xi} + \frac{2\pi Q}{2^n}\;,\qquad p_{E_j,\,\xi} = \braket{E_{j,\,\xi}|V \rho_\xi V^\dagger|E_{j,\,\xi}} \;.
\ee
By algebraic manipulation, Eq.~\eqref{pp2} can also be written as
\be\label{pqsin}
p_{q,\,\xi} = \sum_{j=0}^{d-1} p_{E_j,\,\xi} \left(\frac{1}{2^n}\frac{\sin(2^n \alpha_{j,Q}/2)}{\sin(\alpha_{j,Q}/2)}\right)^2\;.
\ee
In the limit  $n\to \infty$, the probability distribution $p_{q,\,\xi}$ converges to the probability distribution $p_{E_j,\,\xi}$ corresponding to a controlled energy measurement $\mathscr M_{V,\,\xi}$. 

\begin{figure}[h]
\includegraphics[width=0.95\columnwidth]{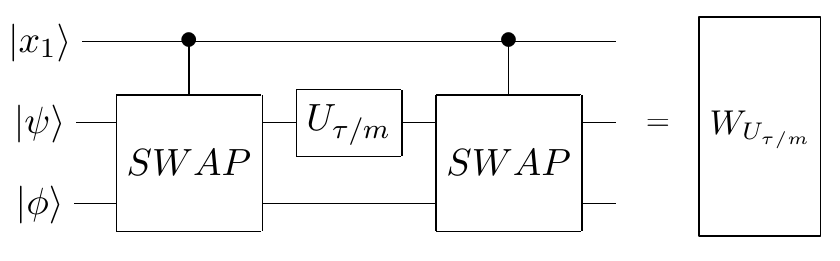}
\caption{Circuit diagram of $W_{U_{\tau/m}}$. \label{w}}
\end{figure}

We now explain how to implement the controlled time-evolution operator without full knowledge of the Hamiltonian. For a more detailed treatment, we refer the reader to Ref.~\cite{nakayama2015quantum}. For notational simplicity consider the case $l=1$, so that the problem is to approximate the action of $\mathcal C_{U_\tau}$ on the state $\rho_1 = \ket{x_1}\bra{y_1} \otimes V \rho_\xi V^\dagger$. Since $\mathcal C_{U_\tau}$ is not actually available, it is replaced by $m$ applications of the superoperator $\Gamma_{U_{\tau/m}}$, constructed as follows. 
First of all, we introduce an ancilla having the same dimension as the main system, so that the total Hilbert space is $\mathcal H_c^{\otimes n} \otimes \mathcal H \otimes \mathcal H_a$, with $\mathcal H_a=\mathbb C^d$. The ancilla is prepared in the maximally mixed state. Therefore, the state of the first control qubit, the main system and the ancilla before application of $\mathcal C_{U_\tau}$ is $\rho'_1 = \ket{x_1}\bra{y_1} \otimes V \rho_\xi V^\dagger \otimes \mathbb I_d/d$. Let us define the following quantum operation,
\be
W_{U_\tau} \eqd C_{SWAP} (\mathbb I_2 \otimes U_\tau \otimes \mathbb I_d)\, C_{SWAP}\;,
\ee 
where $C_{SWAP}$ is the controlled-SWAP gate acting as follows on $\mathcal H_c \otimes \mathcal H \otimes \mathcal H_a$: $C_{SWAP}(\ket 0 \otimes \ket \psi \otimes \ket \phi) = \ket 0 \otimes \ket \phi \otimes \ket \psi$ and $C_{SWAP}(\ket 1 \otimes \ket \psi \otimes \ket \phi) = \ket 0 \otimes \ket \psi \otimes \ket \phi$. The key remark is that implementation of $W_{U_\tau}$ does not require knowledge of the Hamiltonian, but  makes use instead of the uncontrolled version of the time-evolution operator $U_\tau$ (see also Fig. \ref{w}). 
\par
We now subdivide $\tau$ into $m$ subintervals of length $\tau/m$. During each subinterval, $W_{U_{\tau/m}}$ is applied; then the ancilla is discarded; finally, the ancilla is refreshed to its initial state. For instance, after the first interval, one obtains $\Gamma_{U_{\tau/m}}[\rho_1]\otimes \mathbb I_d/d$, where 
\be\label{biggamma}
\Gamma_{U_{\tau/m}}[\rho_1] \eqd \tr_{\mathcal H_a}\left(W_{U_{\tau/m}} \rho_1'\,  W^\dagger_{U_{\tau/m}}\right)\;.
\ee
A simple computation reveals that
\be
\Gamma_{U_{\tau/m}}[\rho_1] = \frac{1}{d}\tr\left(U_{\tau/m}^{y_1-x_1}\right)\, \mathcal C_{U_{\tau/m}}[\rho_1]\;. 
\ee
For future convenience, we write
\be
\frac{1}{d}\,\tr\left(U_{\tau/m}\right) = a_{\tau/m}\, e^{i\phi_{\tau/m}}\;,
\ee
where $a_{\tau/m}\in \mathbb R^+$ and $\phi_{\tau/m}\in \mathbb R$. Note that, since $x_1-y_1 \in \{-1,0,1\}$, one can write
\be
\Gamma_{U_{\tau/m}}^m[\rho_1] = a_{\tau/m}^{|x_1-y_1|m}\, e^{i(y_1-x_1)m\phi_{\tau/m}}\,\mathcal C_{U_{\tau}}[\rho_1]\;.
\ee
Universal controllization basically replaces $\mathcal C_{U_{\tau}}$ with $\Gamma^m_{U_{\tau/m}}$. In the limit $m\to\infty$, it can be proven that the error
\be
\epsilon_m \eqd \left[\tr\left(U_{\tau/m}\right)/d\right]^m-1
\ee
tends to zero.
A realistic controlled energy measurement is thus obtained by substituting each application of $\mathcal C_{U^{2^{l-1}}_{\tau}}$ by $2^{l-1}m$ applications of  $\Gamma_{U_{\tau/m}}$. For instance, instead of Eq.~\eqref{ts1}, one would have
\be
\frac{1}{2^n} \sum_{X,Y=0}^{2^n-1} \pi_{X,Y} e^{i(Y-X)m \phi_{\tau/m}} \ket{x}\!\bra{y} \otimes U_\tau^X V \rho_\xi V^\dagger (U^\dagger_\tau)^Y\,,
\ee 
where we defined
\be
\pi_{X,Y}\eqd \prod_{l=1}^n a_{\tau/m}^{|x_l-y_l| 2^{l-1} m }\;.   
\ee
After applying the inverse quantum Fourier transform and measuring in the computational basis, the probability of obtaining the outcome $q\in \{0,1\}^{\times n}$ is
\be\label{pqfin}
p_{q,\,\xi} = \frac{1}{2^{2n}} \sum_{j=0}^{d-1} p_{E_j,\,\xi} \sum_{X,Y=0}^{2^n-1} \pi_{X,Y}\, e^{i (Y-X) \beta_{j,Q}}\;, 
\ee
with 
\be
\beta_{j,Q} \eqd \alpha_{j,Q} + m \phi_{\tau/m}\;.
\ee
Eq.~\eqref{pqfin} can be further expanded by rewriting it as follows, 
\be\label{pqcos}
\begin{split}
p_{q,\,\xi} = \frac{1}{2^{2n}} \sum_{j=0}^{d-1} p_{E_j,\,\xi} \prod_{l=1}^n \sum_{u,v=0}^1  a_{\tau/m}^{|u-v| 2^{l-1} m} e^{i(v-u) 2^{l-1}\beta_{j,Q}}\\
= \frac{1}{2^n} \sum_{j=0}^{d-1} p_{E_j,\,\xi}  \prod_{l=1}^n \left[1+ a_{\tau/m}^{ 2^{l-1} m}\cos\left(2^{l-1} \beta_{j,Q}\right) \right]
\end{split}
\ee
If $m\to \infty$, then $\phi_{\tau/m}\to 0$ and $a_{\tau/m}\to 1$, so that Eq.~\eqref{pqcos} converges to Eq.~\eqref{pqsin}. Therefore, a realistic controlled energy measurement allows to approximate to any desired precision a controlled energy measurement $\mathscr M_{V,\,\xi}$, without requiring any a priori knowledge about the parameter $\xi$. The result is asymptotic, in the sense that the previous statement holds when both, the number of control qubits $n$ and the number of subintervals $m$, go to infinity. In the next section, we discuss in detail a prototypical example and find that, even for small values of $n$ and $m$, a controlled energy measurement can be well approximated, and thus a precision enhancement is possible compared to the optimal Braunstein-Caves measurement.

\section{Examples: quantum magnetometry}\label{S7}

\subsection{Qubit magnetometry: estimating the direction of a magnetic field}
The problem is to estimate the polar angular direction $\xi$ of an external magnetic field of known magnitude $B$ by use of a qubit probe, with Hilbert space $\mathcal H = \mathbb C^2$ and Hamiltonian  $H_\xi=\omega(\cos\xi\, \sigma_z + \sin\xi\, \sigma_x)$ (the energy splitting $\omega$ is proportional to $B$, thus it is assumed to be known). In the first part of this section, we compare the family of regular measurements $\rsfscr R$ with the non-regular family $\rsfscr E$ of controlled energy measurements. Next, we analyze the problem in a more physical setting, by evaluating the performance achievable via realistic controlled energy measurements. 

The probe is initialized at time $t=0$ in the state $\ket{\psi_0} = \ket 0$. The parameter is encoded unitarily for a time $t$, leading to $\ket{\psi_\xi}= U_t \ket{\psi_0}$, with $U_t \eqd \text{exp}(-i H_\xi t)$. Let us suppose first that  only regular measurements are allowed. Then, the best achievable performance is given by the QFI,
\be\label{qfimag}
\mathcal F_\xi^{(Q)}(\ket{\psi_\xi}) = 4 \sin^2(\omega t) - \sin^2(2 \omega t) \sin^2\xi\;. 
\ee
Optimizing also over the initial preparation $\ket{\psi_0}$, one arrives at the CQFI $\mathcal F_\xi^{(Q,\,C)}\!= 4 \sin^2(\omega t)$.
\par
\begin{figure}[h!]
\includegraphics[width=0.95\columnwidth]{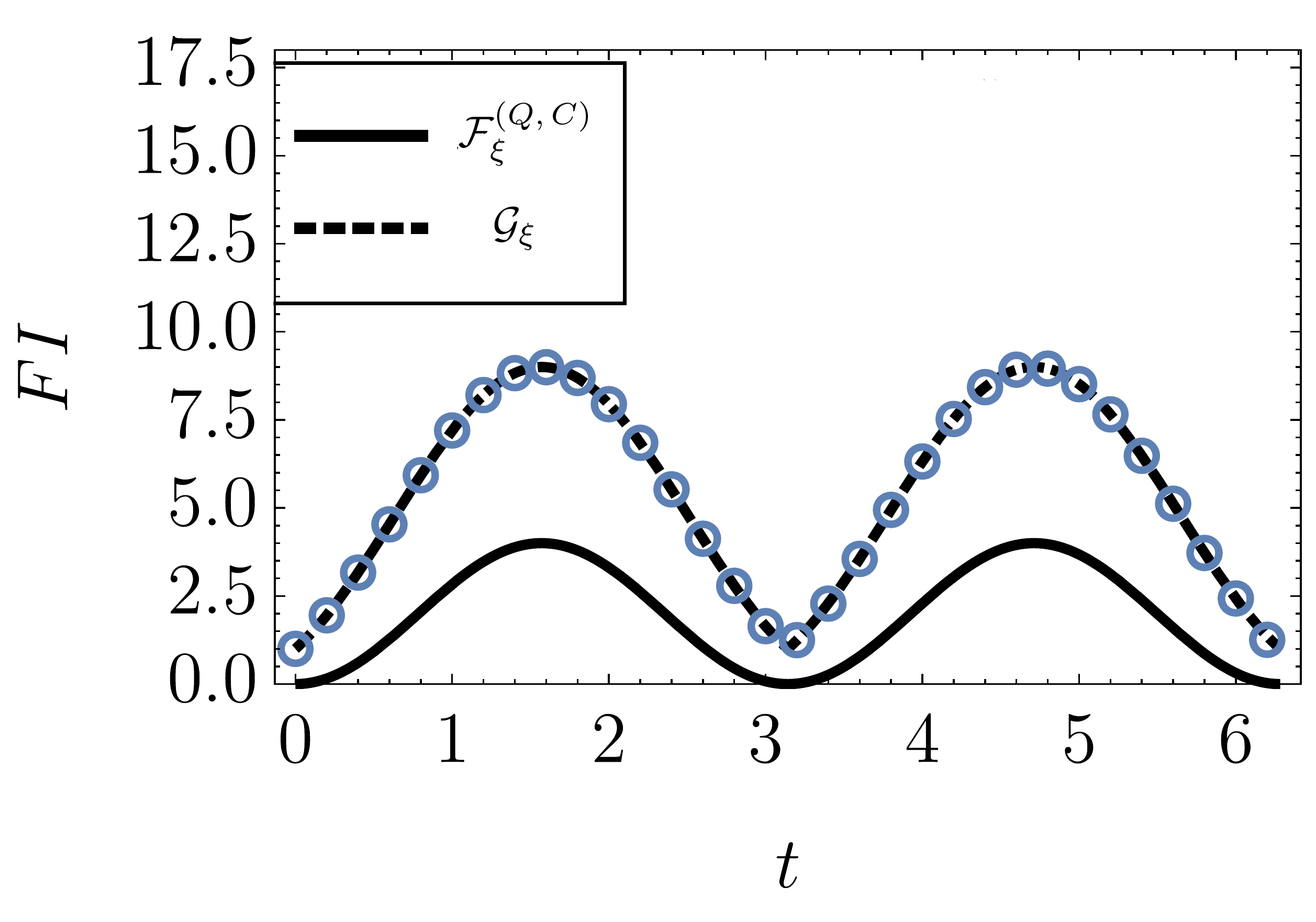}
\caption{Comparison between the optimal Braunstein-Caves measurement and the optimal controlled energy measurement, for the estimation of the polar angular direction of a magnetic field via a qubit probe. The solid line is the CQFI, while the dashed line corresponds to $\mathcal G_\xi$, computed by Eq.~\eqref{gximagf}. The circular marks denote $\mathcal G_\xi$, computed by numerical optimization, from its definition \eqref{cefid}, thus confirming that the bound given in Prop.~\ref{pr1} is saturated.\label{cmp1}}
\end{figure}

Suppose instead that the measurement is taken from the family of controlled energy measurements. Then, the best achievable precision is given by the information quantity $\mathcal G_\xi$ of Eq.~\eqref{cefid}. 
Let us compute the matrix $S_\xi$, built from the eigenvectors of $H_\xi$, and its generator $\mathfrak g_{S,\,\xi}$:
\ben
S_{\xi} = \left(\begin{matrix}
-\text{sgn}\left[\cos\left(\frac{\xi}{2}\right)\right] \sin\left(\frac{\xi}{2}\right)  & \text{sgn}\left[\cos\left(\frac{\xi}{2}\right)\right] \cos\left(\frac{\xi}{2}\right)\\
\text{sgn}\left[\sin\left(\frac{\xi}{2}\right)\right] \cos\left(\frac{\xi}{2}\right)  &  \text{sgn}\left[\sin\left(\frac{\xi}{2}\right)\right] \sin\left(\frac{\xi}{2}\right)
\end{matrix}\right)\;,
\een
\ben
\mathfrak g_{S,\,\xi} = \left(\begin{matrix}
0&-\frac{i}{2}\,\text{sgn}(\sin\xi)\\
\frac{i}{2}\,\text{sgn}(\sin\xi)&0\\
\end{matrix}\right)\;,
\een
where $\text{sgn}(x) = |x|/x$. The extremal eigenvectors of $\mathfrak g_{S,\,\xi}$ are
\ben
\ket{\lambda_1(\mathfrak g_{S,\,\xi})}=\frac{1}{\sqrt 2}\;(-i,\;1)^t\;,\quad \ket{\lambda_2(\mathfrak g_{S,\,\xi})}=\frac{1}{\sqrt 2}\;(i,\;1)^t\;.
\een
Since they are equioriented, by Prop.~\eqref{pr2} $\mathcal G_\xi$ can be computed as
\be\label{gximag}
\mathcal G_\xi = \left(\sigma[\mathfrak g_{U,\,\xi}]+\sigma[\mathfrak g_{S,\,\xi}]\right)^2\;.
\ee 
The explicit expressions for $U_t$ and its generator are 
\begin{align}
U_t = \left(\begin{matrix}
 A &  B\\
B  &  A^*\\
\end{matrix}\right)\,, \qquad 
\mathfrak g_{U,\,\xi} & =
\left(\begin{matrix}
- C & D \\
D^* & C
\end{matrix}\right) \notag
\end{align}
where
\begin{align}
A & = \cos\omega t- i\cos\xi\sin\omega t \notag \\
B & = -i\sin\xi\sin\omega t \notag \\ 
C & = \frac{1}{2}\sin\xi\sin 2\omega \notag \\
D & = \left(\cos\xi\cos \omega t-i\sin \omega t\right)\sin\omega t
\;.
\end{align}
After diagonalizing $\mathfrak g_{U,\,\xi}$ and $\mathfrak g_{S,\,\xi}$, one can compute $\mathcal G_\xi$ via Eq.~\eqref{gximag}, which gives
\be\label{gximagf}
\mathcal G_\xi = \mathcal F_\xi^{(Q,\,C)} + 4 |\sin(\omega t)| +1\;.
\ee
As $\mathcal G_\xi > \mathcal F_\xi^{(Q,\,C)}$, the optimal Braunstein-Caves measurement is outperformed. A comparison is shown in Fig.~\ref{cmp1}.
\par
\begin{figure}[h]
\includegraphics[width=0.68\columnwidth]{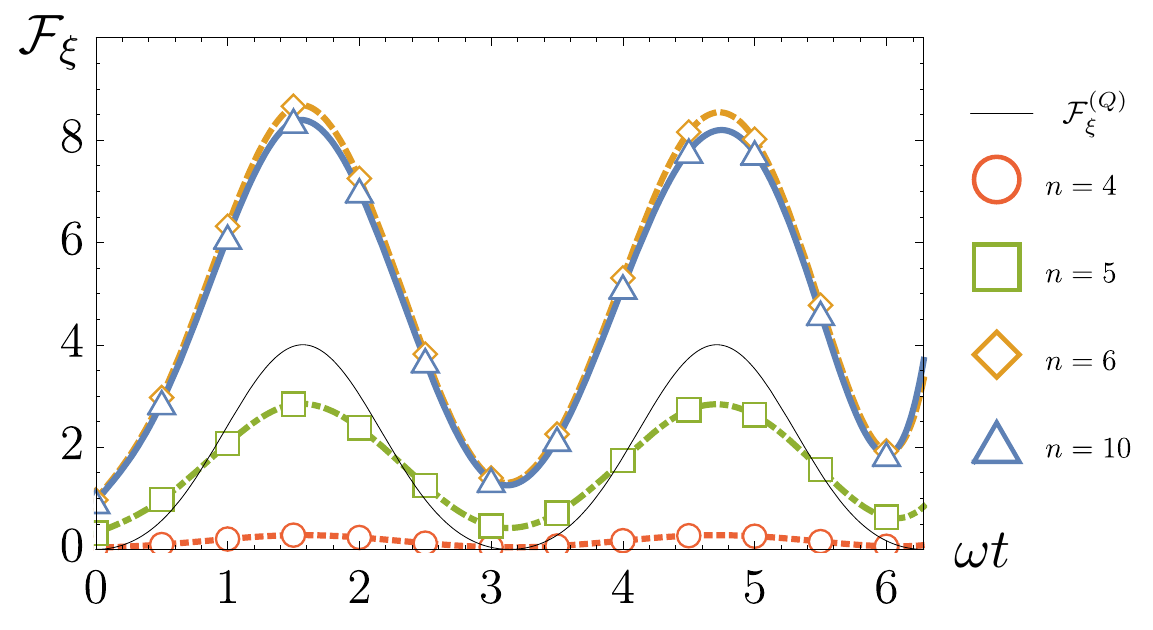}
\includegraphics[width=0.68\columnwidth]{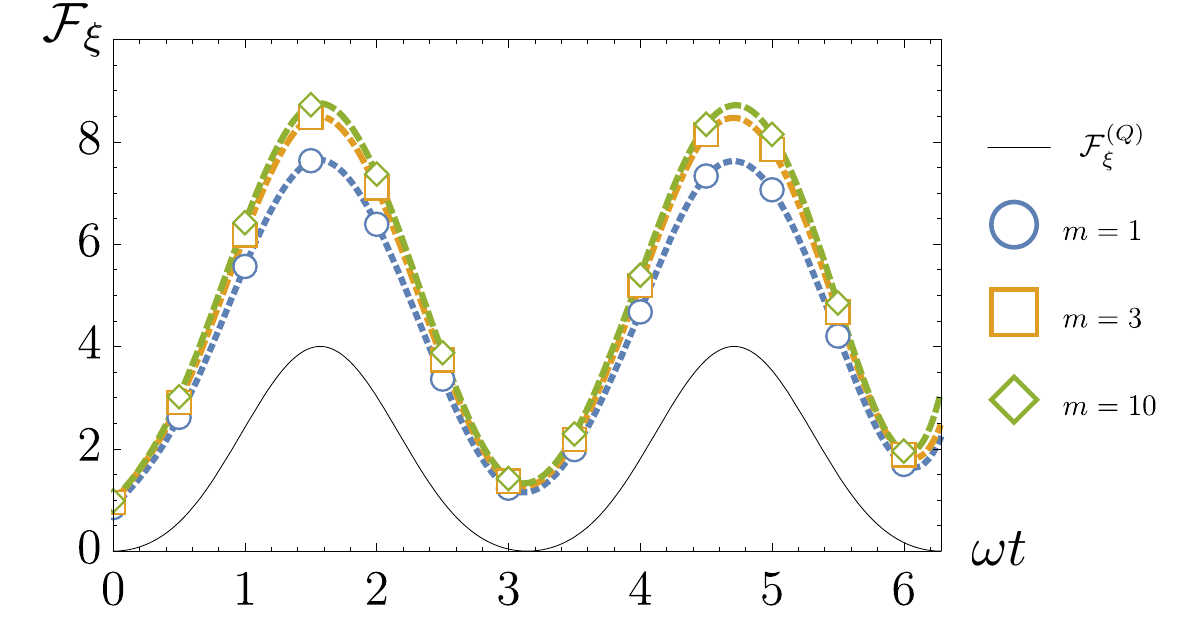}
\caption{ {\emph{Upper panel}: FI of the best-performing realistic controlled energy measurement, for different values of $n$ and fixed $m=5$. Each marker represents the maximum FI, taken over the family of realistic controlled energy measurements for given $n$, $m$, $\tau$ and interrogation time $t$. The curves are obtained by interpolation. The thin solid curve corresponds to the QFI of Eq.~\eqref{qfimag}. Notice that the optimal Braunstein-Caves measurement is outperformed already for $n=6$. \emph{Lower panel}: FI of the best-performing realistic controlled energy measurement, for different values of $m$ and fixed $n=6$. 
Both plots are obtained for $\omega=1$ and $\tau =0.1$ (in the natural units of the problem), while the true value of the parameter is taken to be $\xi=\pi/4$.
\label{fishnm}}}
\end{figure}

Finally, we study numerically the case when the measurement is a realistic controlled energy measurement. This requires to introduce $n$ ancillary qubits and implement the quantum algorithm described in Sect.~\ref{exsec}. In particular, universal controllization is needed to approximate the action of the controlled time-evolution operator $\mathcal C_{U_\tau}$, by subdividing $\tau$ into $m$ subintervals and applying the superoperator $\Gamma_{U_{\tau/m}}$ of Eq.~\eqref{biggamma} in each subinterval. In the limit $n,\,m\to \infty$, one performs the corresponding controlled energy measurement exactly (and thus can achieve $\mathcal G_\xi$ of Eq.~\eqref{gximag}). The two  panels of Fig.~\ref{fishnm} show the performance of the optimal realistic controlled energy measurement, for different values of $n$ and $m$. Reasonably small values of the two parameters (e.g. $n=6$ and $m=3$) are  enough to come close to the ultimate bound $\mathcal G_\xi$ of Eq.~\eqref{gximagf}.
\par
\par
\subsection{Qubit magnetometry: estimating one component of a magnetic field}
\begin{figure}[h]
\flushleft{\quad \includegraphics[width=0.95\columnwidth]{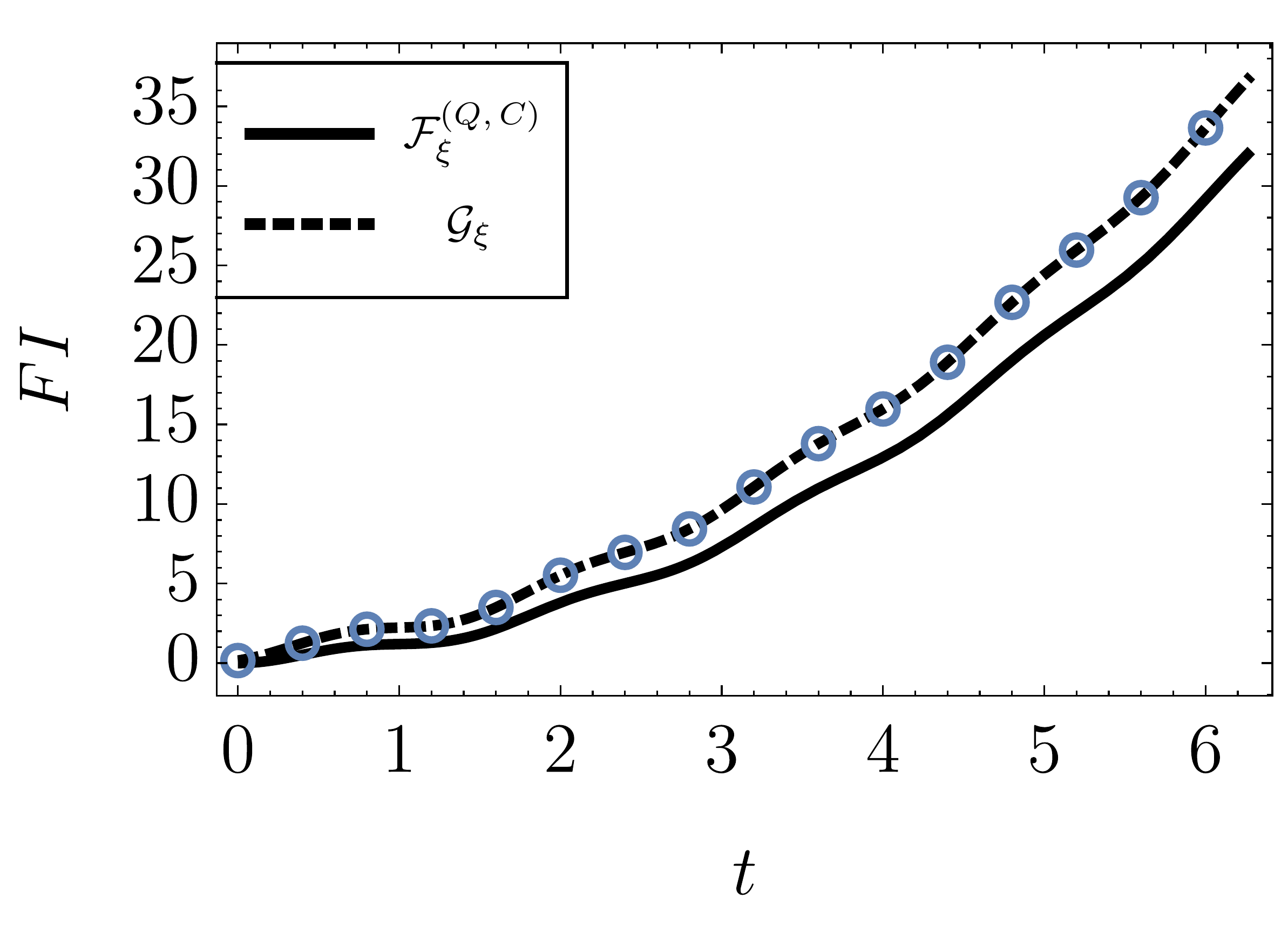}}
\caption{Comparison between the optimal Braunstein-Caves measurement and the optimal controlled energy measurement, for the estimation of one component of a magnetic field via a qubit probe. The solid line is the CQFI, while the dashed line corresponds to $\mathcal G_\xi$. The circular marks denote $\mathcal G_\xi$, computed by numerical optimization, from its definition \eqref{cefid}, thus confirming that the bound given in Prop.~\ref{pr1} is saturated.\label{cmp2}}
\end{figure}

Here the task is to estimate one component of an external magnetic 
field along a given direction (which, without loss of generality, is
 taken to be parallel to the $x$ axis) via a qubit probe. The Hamiltonian is $H_\xi = -\omega \sigma_z + \xi \sigma_x$, with eigenvalues $\pm \Omega_\xi$, where $\Omega_\xi \coloneqq \sqrt{\omega^2+\xi^2}$. As before, one has to compute the relevant matrices $U_t,\,S_\xi$, and their corresponding generators. Concerning $U_t$ and $g_{U,\,\xi}$, we have
\begin{align}
U_t = \left(\begin{matrix}
 A &  B\\
B  &  A^*\\
\end{matrix}\right)\,, \qquad 
\mathfrak g_{U,\,\xi} & =
\left(\begin{matrix}
- C & D \\
D^* & C
\end{matrix}\right) \notag
\end{align}
where
\begin{align}
A & =  \cos (\Omega_\xi t)+\frac{i \omega  
\sin (\Omega_\xi t)}{\Omega_\xi} \\
B & = -\frac{i \xi  \sin (\Omega_\xi t)}{\Omega_\xi} \notag \\ 
C & =  - \frac{ \omega\xi \,  [\sin (2\,\Omega_\xi t)-2\,\Omega_\xi t]}{2\,\Omega_\xi^3} \notag \\
D & = \frac{\sin (2\,\Omega_\xi t) \omega ^2-i \Omega_\xi \cos (2\,\Omega_\xi t) \omega +\Omega_\xi \left(2  t\xi^2+i \omega \right)}{2\,\Omega_\xi^3} \notag
\;.
\end{align} 
$S_\xi$ and its generator $\mathfrak g_{S,\,\xi}$ are instead given by 
\begin{align}
S_{\xi} & =\frac{1}{\sqrt{2\,\Omega_\xi}}
\left(\begin{matrix}
 -\frac{\omega + \Omega_\xi}{\sqrt{\Omega_\xi+\omega}}&\frac{\xi}{\sqrt{\Omega_\xi+\omega}}\\
 \frac{\xi}{\sqrt{\Omega_\xi+\omega}}&\frac{\xi}{\sqrt{\Omega_\xi-\omega}}
\end{matrix}\right)\;, \\
\mathfrak g_{S,\,\xi} & =
\left(\begin{matrix}
 0 & \frac{i \omega }{2 \xi ^2} \\
 -\frac{i \omega }{2 \xi ^2} & 0 \\
\end{matrix}\right)\;.
\end{align}
The CQFI is found by diagonalizing $\mathfrak g_{U,\,\xi}$, i.e.
\be\label{r1}
\mathcal F_\xi^{(Q,\,C)}=\frac{2}{\Omega_\xi^4} \left[2\, \Omega_\xi^2\, t^2\xi^2-\omega ^2 \cos (2\,\Omega_\xi t)+\omega ^2\right]\;.
\ee
Since the eigenvectors of $\mathfrak g_{S,\,\xi}$ are equioriented, $\mathcal G_\xi$ can be computed directly, 
\be\label{r2}
 \mathcal G_\xi=\left(\frac{\omega}{\Omega_\xi^2} + \frac{\sqrt{2 \left[2\,\Omega_\xi^2\, t^2\xi^2-\omega ^2 \cos (2\,\Omega_\xi t)+\omega ^2\right]}}{\Omega_\xi^2} \right)^2\;.
\ee 
A comparison similar to that of Fig.~\ref{cmp1} is shown in Fig.~\ref{cmp2}.
\subsection{NV-center magnetometry}
As a last example, we study the problem of estimating a weak magnetic field via an NV-center in diamond. An NV center consists of a nitrogen atom (N) inside a diamond crystal lattice, having a vacancy (V) in one of its neighboring sites. Two different kinds of the defect are known: the neutral state $NV_0$ and the negatively-charged state $NV_{-}$, which is the most interesting for metrological applications. The $NV_-$ form provides a spin triplet state which can be initialized, manipulated with long coherence time and readout by purely optical means. The reader is referred to the review \cite{rondin2014magnetometry} for more details. 
\par
Neglecting the interactions with the surrounding nuclear spins, and setting $\hbar=1$, the Hamiltonian $H_{NV}\,$ for the triplet state of the NV center can be written in the form
\be
H_{NV} =  \mu\,\bm B \cdot \bm S + D\, S_z^2 + E\, (S_x^2-S_y^2)\;,
\ee
where $\bm B$ is the applied magnetic field and $\bm S =(S_x, S_y, S_z)$ is a vector made up by the three spin 1 matrices
\begin{align}\label{pms1}
S_x =\sqrt 2\, \left(\begin{matrix}
0&1&0\\
1&0&1\\
0&1&0\\
\end{matrix}\right), &  \;\, S_y = \sqrt 2 i\,
\left(\begin{matrix}
0&-1&0\\
1&0&-1\\
0&1&0\\
\end{matrix}\right)\,, \notag \\
 S_z  = 2 &
\left(\begin{matrix}
1&0&0\\
0&0&0\\
0&0&-1\\
\end{matrix}\right)\,.
\end{align} 
Moreover, $D\apeq \pi \,\times\, 1.44\,\si{GHz}$, $E\apeq \pi \,\times\, 50\, \si{kHz}$ and $\mu$ is the Bohr magneton. We work in the weak magnetic field regime, where the transversal components $B_x$ and $B_y$ can be neglected compared to the component $B_z$ aligned along the NV-center defect axis. 

\begin{figure}[h]
\flushleft{\quad
\includegraphics[width=0.95\columnwidth]{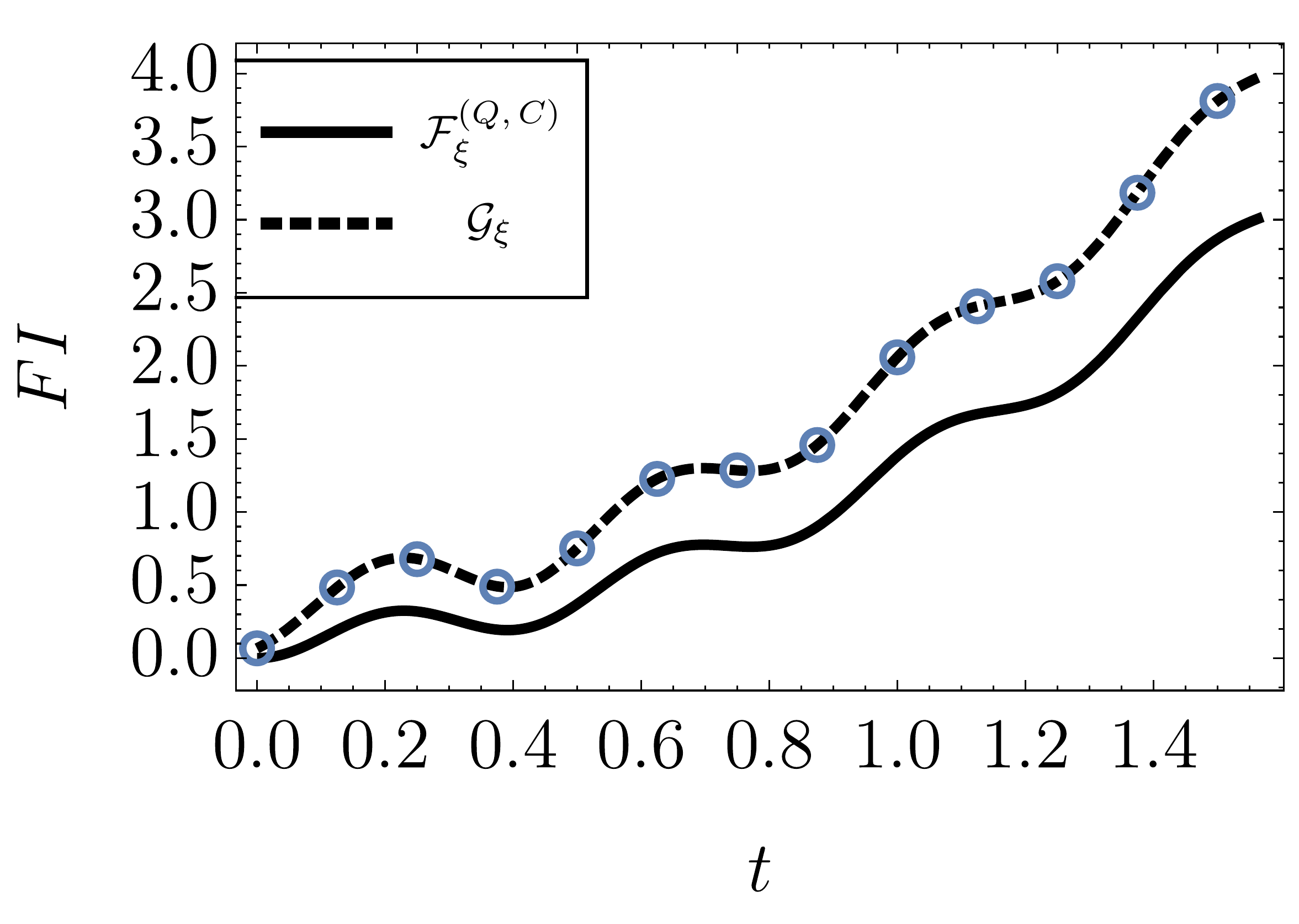}}
\caption{Comparison between the optimal Braunstein-Caves measurement and the optimal controlled energy measurement for the estimation of the magnitude 
of a weak magnetic field via a NV-center in diamond. 
The solid line is the CQFI, while the dashed line corresponds to $\mathcal G_\xi$. The circular marks denote $\mathcal G_\xi$, computed by numerical optimization, from its definition \eqref{cefid}, thus confirming that the bound given in Prop.~\ref{pr1} is saturated.
\label{cmp3}}
\end{figure}

The task is to estimate the field component $B_z$, which from now on we denote conventionally by $\xi$. The CQFI is found to be
\be
\mathcal F_\xi^{(Q,\,C)}=\frac{8  \mu ^2 \left[2\xi^2 \mu^2 t^2 \chi^2+E^2-E^2 \cos \left(4 \chi t\right)\right]}{\chi^4}\;,
\ee
where $\chi\eqd \sqrt{\xi^2\mu^2 +4 E^2}$, while $\mathcal G_\xi$ is given by
\be
\mathcal G_\xi =\! \left(\frac{2 E \mu }{\chi^2}+2\sqrt 2\mu\frac{ \sqrt{2\xi^2 \mu^2 t^2 \chi^2+E^2-E^2 \cos \left(4 \chi t\right)}}{\chi^2} \right)^2\!.
\ee
A comparison is shown in Fig.~\ref{cmp3}. 
\section{Conclusions}\label{consec}
In this paper, the main focus has been on a class of {\em non-regular} quantum measurements, referred to as controlled energy measurements, that are naturally 
available in the estimation of a general Hamiltonian parameter. We have introduced the information quantity $\mathcal G_\xi$, which gives the 
best achievable precision over such a class, and 
provided an upper-bound to it, that can often be saturated in practice. 
We have also discussed a realistic implementation of controlled energy measurements, which makes use of the phase estimation algorithm and 
a quantum subroutine known as universal controllization. Finally, we 
have applied our results to a few prototypical estimation problems and 
found  a precision enhancement with respect to the optimal 
Braunstein-Caves measurement. 
\par
The difficulty, as a matter of principle, of encoding the (unknown) parameter into the measurement apparatus is solved by making use of the time-evolution generated by the system's Hamiltonian as a resource. In this way, the POVM elements formally  acquire an intrinsic dependence on the parameter, which in turn makes an analysis based only on the quantum Fisher information insufficient to capture the ultimate precision bounds. Our results thus show that for Hamiltonian parameters that are not just phase parameters, it is possible to overcome the Cram\'er-Rao bound by feasible detection schemes, opening new avenues to the precise estimation of physical parameters at the quantum frontier.

\section*{Acknowledgments}
This work has been supported by JSPS 
through FY2017 program (grant S17118) and by SERB through the
VAJRA award (grant VJR/2017/000011). The authors thank M. A. C. Rossi, F. Albarelli, T. Giani, and G. Guarnieri for discussion in the early stages
of this work.
\bibliographystyle{apsrev4-1}
\bibliography{refs}


\end{document}